\documentclass[12pt,preprint]{aastex}
\begin{document}

\title{One-point Statistics of the Cosmic Density Field in Real
     and Redshift Spaces with A Multiresolutional Decomposition}

\author{Hu Zhan and Li-Zhi Fang}

\affil{Department of Physics, University of Arizona, Tucson, AZ 85721}

\begin{abstract}

In this paper, we develop a method of performing the one-point
statistics of a perturbed density field with a multiresolutional
decomposition based on the discrete wavelet transform (DWT).
We establish the algorithm of the one-point variable and
its moments in considering the effects of Poisson sampling and
selection function. We also establish the mapping between the DWT
one-point statistics in redshift space and real space, i.e. the 
algorithm for recovering the DWT one-point statistics from the redshift
distortion of bulk velocity, velocity dispersion, and selection
function. Numerical tests on N-body simulation samples show that 
this algorithm works well on scales from a few hundreds to a few 
h$^{-1}$ Mpc for four popular cold dark matter models.

Taking the advantage that the DWT one-point variable is dependent on 
both the scale and the shape (configuration) of decomposition modes, 
one can design estimators of the redshift distortion parameter ($\beta$) 
from combinations of DWT modes. Comparing with conventional $\beta$ 
estimators, such as quadrupole-to-monopole ratio, the DWT $\beta$ 
estimators are scale-decomposed. It is useful to consider scale-dependent 
effects. When the non-linear redshift distortion is not negligible, the 
quadrupole-to-monopole ratio is a function of scale. This estimator would 
not work without adding information about the scale-dependence, such as the 
power-spectrum index or the real-space correlation function of the 
random field. The DWT $\beta$ estimators, however, do not need such extra 
information. Moreover, the scale-decomposed $\beta$ estimators would also 
be able to reveal the scale-dependence of the bias parameter of galaxies. 
Numerical tests show that the proposed DWT estimators are able to
determine $\beta$ robustly with less than 15\% uncertainty in the 
redshift range $0 \leq z \leq  3$. 

\end{abstract}


\section{Introduction}

The one-point probability distribution function (PDF) of a random mass
density
field $\rho({\bf x})$, or the counts-in-cells (CiC) statistics of a discrete
random distribution such as galaxies, is probably the first statistics
used to reveal the clustering feature of galaxy distribution\footnote{Since
we study only the one-point PDF in this paper, not N-point PDF, hereafter,
PDF stands only for the one-point statistics.}. In the famous work by Edwin
Hubble (1934), he showed that the frequency distribution of galaxy count
$N$ in angular cells is not Gaussian, but lognormal. This result indicates
that the PDF of galaxy distribution is fundamental in characterizing
the cosmic mass and velocity fields.

Although current samples applied for large-scale structure study have much
deeper redshift and much wider angular size than available in Hubble's time,
the one-point statistics is still frequently applied.
This is because the one-point distribution and its moments
contain complete information of the field, which might not be easily detected 
by other conventional
methods. Unlike the Fourier amplitude, CiC is not subject to
the central limit theorem. It can detect the non-Gaussianity of a field
consisting of randomly distributed non-Gaussian clumps, while the PDF of the
Fourier amplitudes is still Gaussian due to the central limit theorem
(Fan and Bardeen 1995). Even the 2nd moment of one-point distribution is
different from the 2nd moment of the Fourier decomposition -- power
spectrum. The
former contains perturbations on scales larger than the size
of the observed sample, while the latter does not. Therefore, one-point
statistics
is applied on various samples of large-scale structures, including
galaxy surveys (e.g. Hamilton 1985; Alimi et al 1990; Gazta\~{n}aga
1992; Szapudi et al 1996; Kim \& Strauss 1998), transmitted
flux of quasars' Ly$\alpha$ absorption spectrum
(e.g. Meiksin \& Bouchet 1995; Gaztan\~{n}aga \& Croft 1999; Zhan \& Fang
2002),
and N-body
simulation samples (e.g. Coles \&  Jones 1991; Taylor \& Watts 2000).

One-point statistics of a density fluctuation field $\delta({\bf x})$ is
given by the distribution of the one-point variable $\delta_R({\bf x}_0)$,
which is a sampling of the field by a window function $W_R({\bf x-x}_0)$
around position ${\bf x}_0$ and on scale $R$, i.e.
\begin{equation}
\delta_R({\bf x}_0)=\int W_R({\bf x-x}_0)\delta({\bf x}) d{\bf x}.
\end{equation}
The variable $\delta_R({\bf x}_0)$ is actually a mean value of the field
at position ${\bf x}_0$ and on scale $R$. The distribution of
$\delta_R({\bf x}_0)$ gives a PDF description of the field.

Eq.(1) is a space(${\bf x}_0$)-scale($R$) decomposition of the field
$\delta({\bf x})$ with bases $W_R({\bf x}_0)$. Most popular windows
$W_R({\bf x-x}_0)$ are Gaussian and top-hat filters. The Gaussian
windows generally are not orthogonal, i.e. the function $W_R({\bf x-x}_0)$
does not satisfy  $\int W_R({\bf x-x}_1) W_R({\bf x-x}_2)d{\bf x} = 0$,
if $\bf{ x}_1 \neq {\bf  x}_2$. Thus, the variables  $\delta_R({\bf x}_0)$
are either incomplete or redundant. In turn, these may lead to 1) loss of
information of the field  $\delta({\bf x})$; and 2) false correlation.
It is possible to construct an orthogonal and complete set of bases
from top-hat windows.
However, they are not localized in the Fourier space. The index $R$
does not refer to a well defined scale $k$. As a consequence, the one-point
statistics with conventional windows are not suitable for problems
with a scale-dependence. For instance, the redshift
distortion on the one-point statistics can be properly estimated only
if we know how the redshift distortion depends on the scale and
shape of the window function. Therefore, a better algorithm for the one-point
statistics is needed, for example, to recover the real-space rms density
fluctuation $\sigma_8$ from redshift distortion.

In this paper, we show that these problems can be solved with a
multiresolutional
decomposition via the discrete wavelet
transform (DWT). The bases of the DWT decomposition are complete,
orthogonal,
and localized in both physical and Fourier spaces. It was shown in the
last few years that the DWT can be employed as an alternative representation
in most conventional statistics of cosmic mass and velocity fields,
including
power spectrum (Pando \& Fang 1998; Fang \& Feng 2000; Yang et al 2001a,
2002), high order correlations (Pando \& Fang 1998; Pando,
Feng \& Fang 2001; Feng, Pando \& Fang 2001), bulk and pair-wise velocity
(Zhan \& Fang 2002; Yang et al. 2001b), and identification of halos and
clusters (Xu, Fang \& Wu 2000). The DWT representation is also
able to reveal statistical features, which might not be easily detected
without a set of decomposition bases localized in both physical and scale
spaces. With the DWT
analysis, the intermittency of the fluctuations of quasars' Ly$\alpha$
transmitted flux has been detected (Jamkhedkar, Zhan \& Fang 2000;
Pando, Feng \& Fang 2000; Zhan, Jamkhedkar \& Fang 2001; Pando et al.
2002).
This property cannot be simply detected by the popular $W_R({\bf x-x}_0)$
or Fourier decompositions.

We also show that the DWT decomposition is very useful for one-point
statistics. We establish the algorithm of the one-point variable and
its moments in considering various corrections, such as Poisson sampling and
selection function. To demonstrate the advantage of the DWT one-point
statistics, we show that one can map between the one-point
statistics in real and redshift spaces scale-by-scale, taking account
the distortion due to
bulk velocity, velocity dispersion and selection function.
With these results one can construct $\beta$ (redshift distortion parameter)
estimators with moments of the DWT one-point variables. Unlike
conventional estimators, the DWT estimators do not need
extra-information or {\it ad hoc} assumption when the non-linear redshift
distortion is not negligible.

This paper is organized as follows. \S 2 presents the algorithm for
the PDF of one-point variables with the DWT space-scale decomposition.
\S 3 discusses the second moment of the one-point variables in considering
the effects of Poisson sampling and selection function. In \S 4 we develop
the
theory of redshift distortion of the DWT one-point variables. \S 5
tests the theoretical results in \S 4 on N-body simulation samples.
The emphases are the redshift-to-real-space
mapping of the diagonal and off-diagonal second moments. \S 6 shows the
application in estimating the redshift distortion parameter. Finally, the
conclusions and discussions are given in \S 7. The Appendix provides
relevant formulae for quantities defined in \S 4. We release the codes for
calculating these quantities via anonymous ftp at
\url{samuri.la.asu.edu/pub/zhan/DWTCiC.tgz}.

\section{One-point statistics in the DWT representation}

\subsection{DWT variable of one-point statistics}

As emphasized in \S 1, a key problem for one-point statistics is to
have proper window functions $W_R({\bf x})$, and define the variable
with eq.(1). In the DWT analysis, this is done by the so-called scaling
function. Let us first briefly introduce the DWT-decomposition of a
random field. For the details of the mathematical properties of the
DWT see Mallat (1989a,b); Meyer (1992); Daubechies (1992), and
for physical applications see Fang \& Thews (1998).

To simplify the notation, we first consider a
one-dimensional (1-D) density fluctuation $\delta(x)$ on a spatial range
from $x=0$ to $L$. We first divide the space $L$ into $2^j$ segments
labeled by $l=0,1,...2^j-1$. Each segment is of size $L/2^j$. The index
$j$ can be a positive integer. It stands for a length scale $L/2^j$.
The higher the $j$, the smaller the length scale. Even though
we often refer some properties to a (range of) $j$ in the analyses
below, it must be read as an association of the properties with the
length scale $L/2^j$. The index $l$ is for
position, and it corresponds to the spatial range $lL/2^j < x < (l+1)L/2^j$.
That is, the space $L$ is decomposed into cells $(j,l)$.

In the DWT analysis, each cell $(j,l)$ supports two compact functions:
the scaling function $\phi_{j,l}(x)$ and the wavelet $\psi_{j,l}(x)$.
One example of such functions is the Daubechies 4 (D4) 
wavelets\footnote{Unless mentioned in contrast to the scaling functions, 
wavelets, as analysis tools, always involve the scaling functions.}, 
which have to be constructed recursively (Daubechies 1992; see also 
Fang \& Thews 1998). The properties of D4 wavelets are listed below and 
in the appendix \S A.1.
We show, in Fig. 1, examples of the D4 scaling function
and wavelet, and their Fourier transform $\hat{\phi}_{j,l}(k)$ and
$\hat{\psi}_{j,l}(k)$. One can see from Fig. 1 that both the scaling
function and the wavelet are localized in Fourier space as well as in 
physical space. Generally,
$\phi_{j,l}(x)$ and $\psi_{j,l}(x)$ are localized in cell $(j,l)$,
$\hat{\phi}_{j,l}(k)$ is localized in the scale range
$|k|\leq 2\pi 2^j/L$, and $\hat{\psi}_{j,l}(k)$ in the range $k\pm k/2$,
where $|k|=2\pi 2^j/L$.

Obviously, the scaling function $\phi_{j,l}(x)$ is a window function
on scale $j$ and around the segment $l$. It can be used to measure the mean
field in cell $(j,l)$
\begin{equation}
\delta_{j,l}=\frac{\int_{0}^{L} \delta(x)\phi_{j,l}(x)dx}
   {\int_{0}^{L} \phi_{j,l}(x)dx}=
   \frac {1}{\int_{0}^{L} \phi_{j,l}(x)dx} \epsilon_{j,l},
\end{equation}
where $\epsilon_{j,l}$ is called scaling function coefficient (SFC),
given by
\begin{equation}
\epsilon_{j,l}= \int_{0}^{L} \delta(x)\phi_{j,l}(x)dx.
\end{equation}
In analogous to eq.(1), eqs.(2) and (3) show that the SFC $\epsilon_{j,l}$
or
$\delta_{jl}$ can be employed as the variable for one-point
statistics.

Fig. 1 also demonstrates that wavelets are admissible, i.e.
$\int \psi_{j,l}(x)dx=0$, and therefore, they are used to measure the
fluctuations of a field with respect to its mean in cells $(j,l)$
\begin{equation}
\tilde{\epsilon}_{j,l}=\int \delta(x)\psi_{j,l}(x)dx,
\end{equation}
where $\tilde{\epsilon}_{j,l}$ is called the wavelet function
coefficient (WFC).

The scaling function $\phi_{j,l}(x)$ and the wavelet
$\psi_{j,l}(x)$ satisfy a set of orthonormal relations as
\begin{eqnarray} \label{eq:1D-phi-orth}
\int \phi_{j,l}(x)\phi_{j,l'}(x)dx &=& \delta^K_{l,l'},\\
\label{eq:1D-psi-orth}
\int \psi_{j,l}(x)\psi_{j',l'}(x)dx &=& \delta^K_{j,j'} \delta^K_{l,l'},\\
\label{eq:1D-phi-psi-orth}
\int\phi_{j,l}(x)\psi_{j',l'}(x)dx &=& 0, \qquad \mbox{if $j'\geq j$},
\end{eqnarray}
where $\delta^K_{m,n}$ is the Kronecker delta function.
To be consistent with eq.(5), the scaling function $\phi_{j,l}(x)$
is normalized to
\begin{equation}
\int \phi_{j,l}(x)dx= \sqrt{\frac{L}{2^j}}.
\end{equation}
With these properties, a 1-D density field $\delta(x)$ can be
decomposed into
\begin{equation}
\delta(x) =
  \sum_{l=0}^{2^j-1}\epsilon_{j,l}\phi_{j,l}(x) +
  \sum_{j'=j}^{\infty} \sum_{l=0}^{2^{j'}-1}
   \tilde{\epsilon}_{j',l} \psi_{j',l}(x).
\end{equation}
The second term on the r.h.s of eq.(9) contains only the information of
the field $\delta(x)$ on scales equal to and less than $L/2^j$.
Therefore, the $2^j$ variables (SFCs) $\epsilon_{j,l}$, with
$l=0...2^j-1$, completely describe the behavior of the field on scales 
larger than $L/2^j$. For instance, if the resolution
of the sample is $L/2^J$, $\delta(x)$ can be described by the $2^J$ SFCs
$\epsilon_{J,l}$. Because of the orthogonal relation eqs.(5) and (7),
the $2^j$ variables (SFCs) $\epsilon_{j,l}$ are independent and
irredundant.

In a functional space consisting of functions, which are smoothed
on scales equal to and less than $L/2^j$, the completeness of the scaling
function  $\phi_{j,l}(x)$ can be expressed as
\begin{equation}
\sum_{l=0}^{2^j-1}\phi_{j,l}(x)\phi_{j,l}(x') = \delta^D(x-x'),
\end{equation}
where $\delta^D(x-x')$ is the Dirac Delta function.

All the above-mentioned results can be easily generalized to
three-dimensional
(3-D) fields. Let us consider a 3-D distribution $\delta({\bf x})$ in a
volume
${\bf x}=(0,0,0)$ to $(L_1,L_2,L_3)$. Similar to the 1-D case, we divide
the volume $L_1\times L_2 \times L_3$ into cells $({\bf j,l})$, where
${\bf j}=(j_1,j_2,j_3)$ refers to the length scale
$(L_1/2^{j_1}, L_2/2^{j_2}, L_3/2^{j_3})$, and ${\bf l}=(l_1,l_2,l_3)$
refers to the spatial range of the cell,
$l_iL_i/2^{j_i} < x_i \leq (l_i+1)L_i/2^{j_i}$ and  $l_i=0...2^{j_i}-1$,
where $i=1,2,3$. The one-point statistical variable of the field
$\delta({\bf x})$ is
\begin{equation}
\epsilon_{\bf j,l}=\int \delta({\bf x})\phi_{\bf j,l}({\bf x})d{\bf x},
\end{equation}
where the 3-D scaling function (3-D window) $\phi_{\bf j,l}({\bf x})$
is given by a direct product of 1-D scaling functions as
\begin{equation}
 \phi_{\bf j,l}({\bf x})=\phi_{j_1,l_1}(x_1)
  \phi_{j_2,l_2}(x_2)\phi_{j_3,l_3}(x_3).
\end{equation}

One can also construct 3-D wavelets $\psi_{\bf j,l}({\bf x})$ as a direct
product of 1-D scaling functions and wavelets (see Yang et al 2002), 
and generalize eq.(9)
into 3-D as
\begin{equation}
\delta({\bf x}) =
  \sum_{l_1=0}^{2^{j_1}-1}\sum_{l_2=0}^{2^{j_2}-1}
   \sum_{l_3=0}^{2^{j_3}-1}
  \epsilon_{\bf j,l}\phi_{\bf j,l}({\bf x}) +
  \mbox{terms of $\psi_{\bf j',l'}$ with  $j'_i\geq j_i$}.
\end{equation}
Since the second term on the r.h.s. is not needed in calculating
the one-point statistics, therefore, we do not show the
details of $\psi_{\bf j',l'}$, but only give the orthogonal
relations
\begin{equation}
\int\phi_{\bf j,l}({\bf x})\psi_{\bf j',l'}({\bf x})
d{\bf x} =0, \ \ \ \mbox{if $j'_i\geq j_i$, $i=1,2,3$}.
\end{equation}
If we consider only functions smoothed on scales equal to
and less than the scale of ${\bf j}$ in each dimension, the set of bases 
$\phi_{\bf j,l}({\bf x})$ is orthogonal and complete.

\subsection{Moments of DWT one-point distribution}

In one-point statistics, the statistical feature of the field
$\delta({\bf x})$ is characterized by the moments of the distribution
of $\epsilon_{\bf j,l}$. The $m$th moment is
$\langle \epsilon_{\bf j,l}^m \rangle$, where $\langle...\rangle$
stands for an ensemble average. Using eq.(11), we have
\begin{equation}
\langle \epsilon_{\bf j,l}^m \rangle =
 \int \langle \delta({\bf x}_1)...\delta({\bf x}_m) \rangle
    \phi_{\bf j,l}({\bf x}_1)...\phi_{\bf j,l}({\bf x}_m)
    d{\bf x}_1...d{\bf x}_m.
\end{equation}
For a homogeneous field, these moments are ${\bf l}$-independent. Moreover,
when the ``fair sample hypothesis" (Peebles 1980) holds, or
equivalently, when the random field is ergodic, the $2^{j_1+j_2+j_3}$ SFCs
$\epsilon_{\bf j,l}$, $l_i=0...2^{j_i}-1$, $i=1,2,3$ for a given
scale ${\bf j}$ can be considered as $2^{j_1+j_2+j_3}$ independent
measurements, because they are measured by projecting onto the mutually
orthogonal basis $\phi_{\bf j,l}({\bf x})$. Accordingly, the
$2^{j_1+j_2+j_3}$ SFCs form a statistical ensemble on the scale ${\bf j}$.
This ensemble represents actually the one-point distribution of
$\epsilon_{\bf j,l}$ over the DWT modes at a given scale ${\bf j}$.
Thus, the average over ${\bf l}$ is a fair estimation of the ensemble
average, i.e.
\begin{equation}
\langle \epsilon_{\bf j,l}^m \rangle =\frac{1}{2^{j_1+j_2+j_3}}
\sum_{l_1=0}^{2^{j_1}-1}\sum_{l_2=0}^{2^{j_2}-1}\sum_{l_3=0}^{2^{j_3}-1}
\epsilon_{\bf j,l}^m.
\end{equation}

We now consider the 2nd moment $\langle \epsilon_{\bf j,l}^2 \rangle$.
 From eq.(15), the 2nd moment is determined by the usual two point
correlation function
$\langle \delta({\bf x}_1)\delta({\bf x}_2) \rangle$.
We define a dimensionless 2nd moment on scale ${\bf j}$ by
\begin{equation}
D_{\bf j} \equiv \frac{2^{(j_1+j_2+j_3)}}{L_1L_2L_3}
\langle \epsilon_{\bf j,l}^2 \rangle =
\frac{1}{L_1L_2L_3}
\sum_{l_1=0}^{2^{j_1}-1} \sum_{l_2=0}^{2^{j_2}-1}\sum_{l_3=0}^{2^{j_3}-1}
\epsilon^2_{\bf j,l}.
\end{equation}
Using eqs.(12) and (15), one can rewrite eq.(17) as
\begin{equation}
D_{\bf j} = \frac{1}{L_1L_2L_3}
  \sum_{n_1 = - \infty}^{\infty}
  \sum_{n_2 = - \infty}^{\infty}
  \sum_{n_3 = - \infty}^{\infty}
  |\hat{\phi}(n_1/2^{j_1})\hat{\phi}(n_2/2^{j_2})
  \hat{\phi}(n_3/2^{j_3})|^2 P(n_1,n_2,n_3),
\end{equation}
where $\hat{\phi}(n)$ is the Fourier transform of the basic scaling function
(see eq.(A1) in Appendix A). The term $P(n_1,n_2,n_3)$ in eq. (18) is
the Fourier power spectrum defined by
\begin{equation}
P(n_1,n_2,n_3)=
 \langle\hat{\delta}({\bf k})\hat{\delta}^{\dagger}({\bf k})\rangle,
\end{equation}
where ${\bf k}=(k_1,k_2,k_3)$ with $k_i=2\pi n_i/L_i$, and
\begin{equation}
\hat{\delta}({\bf k}) =\frac{1}{L_1L_2L_3}\int_0^{L_1}\int_0^{L_2}
\int_0^{L_3}
\delta({\bf x})e^{-i{\bf k}\cdot{\bf x}}d{\bf x}.
\end{equation}

One can compare the 2nd moment $D_{\bf j}$ with the DWT power spectrum
given by Fang \& Feng (2000) as
\begin{equation}
P_{\bf j} = \frac{1}{2^{j_1+j_2+j_3}}
  \sum_{n_1 = - \infty}^{\infty}
  \sum_{n_2 = - \infty}^{\infty}
  \sum_{n_3 = - \infty}^{\infty}
  |\hat{\psi}(n_1/2^{j_1})\hat{\psi}(n_2/2^{j_2})
  \hat{\psi}(n_3/2^{j_3})|^2 P(n_1,n_2,n_3),
\end{equation}
where $\hat{\psi}(n)$ is the Fourier transform of the basic wavelet.
Therefore, both the DWT power spectrum and the DWT one-point statistics
can be calculated with a single DWT decomposition. The former relies on the
wavelet, while the latter on the scaling function.

The Fourier transform of the scaling function $\hat{\phi}(n)$ is
non-zero in $n$-space where $|n| \leq 1$, and  $\hat{\psi}(n)$ is mainly
in $1/2 < |n| <1$ (Fig. 1 and eq.(A4)). Therefore, the DWT power spectrum
$P_j$
is actually a banded Fourier power spectrum in the wavenumber range
$\pi 2^{j_i}/L_i < |k_i| < \pi 2^{j_i+1}/L_i$, while the 2nd moment
$D_j$ contains all powers with $|k_i| \leq \pi 2^{j_i+1}/L_i$.
It is well known that the 2nd moment of one-point
statistics, such as $D_{\bf j}$ or $\sigma_8$, is sensitive to long
wavelength  behavior of the perturbations. This can also
be seen from the relation between the scaling functions and wavelets
as follows
\begin{equation}
|\hat{\phi}_{j,0}(k)|^2=\frac{1}{2}|\hat{\psi}_{j-1,0}(k)|^2
   + \frac{1}{2} |\hat{\phi}_{j-1,0}(k)|^2 = \sum_{n=1}^{j}
  \frac{1}{2^n}|\hat{\psi}_{j-n,0}(k)|^2 +
   \frac{1}{2^j} |\hat{\phi}_{0,0}(k)|^2.
\end{equation}
That is, the factor $|\hat{\phi}_{j,0}(k)|^2$ in eq.(22) extracts
all powers on scales larger than $L/2^j$. Eq.(22) holds only for
compactly supported discrete wavelets such as Daubechies wavelets.

\section{One-point statistics of galaxy distribution in the DWT
   representation}

\subsection{Galaxy distribution}

We now consider distributions of discrete objects, such as
simulation particles and observed or mock galaxies.
If the position measurement of particles or galaxies is perfectly
precise, the number density distribution of these samples can be
written as
\begin{equation}
n^g({\bf x})=\sum_{m=1}^{N_g}w_m\delta^D({\bf x-x}_m)=
\bar{n}({\bf x})[1+\delta({\bf x})],
\end{equation}
where $N_g$ is the total number of the particles or galaxies, ${\bf x}_m$
is the position of the $m$th galaxy, $w_m$ is its weight, and
$\bar{n}({\bf x})$ is the selection function, which is given by the mean
number density of galaxies when galaxy clustering is absent, and
$\delta({\bf x})$ is the density fluctuation in the distribution.

One can subject $n^g({\bf x})$ to a DWT decomposition. Similar to eq.(13),
we have
\begin{equation}
n^g({\bf x}) =
  \sum_{l_1=0}^{2^{j_1}-1}\sum_{l_2=0}^{2^{j_2}-1}
   \sum_{l_3=0}^{2^{j_3}-1}
  \epsilon^g_{\bf j,l}\phi_{\bf j,l}({\bf x}) +
  \mbox{terms of $\psi_{\bf j',l'}$ with  $j'_i\geq j_i$},
\end{equation}
where
\begin{equation}
\epsilon^g_{\bf j,l}=
\int n^g({\bf x})\phi_{\bf j,l}({\bf x})d{\bf x}
=\sum_{m=1}^{N_g}w_m \phi_{\bf j,l}({\bf x}_m).
\end{equation}

The DWT one-point variables of $\delta({\bf x})$
is now given by
\begin{equation}
\epsilon_{\bf j,l} =
   \int \delta({\bf x})\phi_{\bf j,l}({\bf x})d{\bf x}
  = \int \left [ \frac{n^g({\bf x})}{\bar{n}({\bf x})}-1 \right ]
\phi_{\bf j,l}({\bf x}) d{\bf x}.
\end{equation}
Since the selection function varies slowly, we have
\begin{equation}
\int \frac{n^g({\bf x})}{\bar{n}({\bf x})}\phi_{\bf j,l}({\bf x})d{\bf x}
  \simeq \frac{1}{\bar{n}_{\bf j,l}}
  \int n^g({\bf x})\phi_{\bf j,l}({\bf x})d{\bf x}=
   \frac{\epsilon^g_{\bf j,l}}{\bar{n}_{\bf j,l}},
\end{equation}
where $\bar{n}_{\bf j,l}$ is the mean of the selection faction in the cell
$({\bf j,l})$. The algorithm for $\bar{n}_{\bf j,l}$ is given in
next subsection. Substituting eq.(27) into eq.(26) we have
\begin{equation}
\epsilon_{\bf j,l} \simeq \frac{\epsilon_{\bf j,l}^g}{\bar{n}_{\bf j,l}}
   - \sqrt{\frac{L_1L_2L_3}{2^{j_1+j_2+j_3}}}.
\end{equation}
The second term on the r.h.s. is due the normalization of the scaling
function eq.(8).

\subsection{Selection function in the DWT representation}

By definition of equation (23), selection function $\bar{n}({\bf x})$
is the galaxy distribution if galaxy clustering $\delta({\bf x})$
is absent. In the plane-parallel approximation, selection function
depends only on $x_3$, i.e. the coordinate in the redshift direction
or the line-of-sight (LOS). Thus,
from equation (23) the selection function $\bar{n}(x_3)$ can be
approximated by an average of $n^g({\bf x})$ over the plane
$(x_1, x_2)$, which depends upon the geometry of a real survey.
In a simple case, for example, a mock survey in a simulation box,
it reads
\begin{equation}
\bar{n}(x_3)=\frac{1}{L_1L_2} \int_0^{L_1}\int_{0}^{L_2}
  n^g({\bf x})dx_1 dx_2.
\end{equation}
With eq.(24), eq.(29) yields
\begin{equation}
\bar{n}(x_3) =
\sum_{l_3=0}^{2^{j_3}-1}
    \epsilon^{g}_{00j_3, 00l_3}\phi_{j_3,l_3}(x_3).
\end{equation}

By definition, $\bar{n}_{\bf j,l}$ is the mean of $\bar{n}(x_3)$ in
the cell $({\bf j,l})$. Using eq. (30), we have
\begin{equation}
\bar{n}_{\bf j,l} =\sqrt{\frac{2^{j_1+j_2+j_3}}{L_1L_2L_3}}
    \int \bar{n}(x_3) \phi_{j,l}({\bf x}) d{\bf x}
   = \sqrt{\frac{2^{j_3}}{L_3}}\epsilon^{g}_{00j_3, 00l_3}.
\end{equation}
That is, the selection function can be approximately expressed by
the SFC of the (observable) galaxy distribution $n^g({\bf x})$.

\subsection{Effect of Poisson sampling}

The observed galaxy distributions $n^g({\bf x})$ are considered to be a
Poisson sampling with an intensity
$n({\bf x})=\bar{n}({\bf x})[1+ \delta({\bf x})]$.
In this case, the characteristic function of the galaxy
distribution $n^g({\bf x})$ is
\begin{equation}
\langle [e^{i\int n^g({\bf x})u({\bf x})d{\bf x}}] \rangle_P=
\exp\left \{ \int d{\bf x} n({\bf x})[e^{iu({\bf x})}-1] \right \},
\end{equation}
where $\langle ...\rangle_P$ is the average for the Poisson
sampling. The $m$-point correlation functions of $n^g({\bf x})$ are
given by
\begin{equation}
\langle n^g({\bf x}_1)...n^g({\bf x}_m)\rangle_P =\frac{1}{i^m} \left [
\frac {\delta^m Z}{\delta u({\bf x}_1)... \delta u({\bf x}_m)}
  \right ]_{u=0}.
\end{equation}
We have then
\begin{equation}
\langle n^g({\bf x})\rangle_P = n({\bf x}),
\end{equation}
and
\begin{equation}
\langle n^g({\bf x})n^g({\bf x}')\rangle_P = n({\bf x})n({\bf x}') +
\delta^D({\bf x}-{\bf x}')n({\bf x}).
\end{equation}
Substituting $n({\bf x})=\bar{n}({\bf x})[1+ \delta({\bf x})]$ into
eq.(35), we have
\begin{equation}
\langle \delta({\bf x})\delta({\bf x'})\rangle=\left \langle
\frac{\langle n^g({\bf x})n^g({\bf x}')\rangle_P}
  {\bar{n}({\bf x})\bar{n}({\bf x}') } \right \rangle
- \left \langle \delta^D({\bf x}-{\bf x}')\frac{1}{\bar{n}({\bf x})}
     \right \rangle -1.
\end{equation}
Subjecting eq.(36) to a scaling function projection, and using eq.(27),
we have approximately
\begin{equation}
\langle \epsilon_{\bf j,l}\epsilon_{\bf j',l'}\rangle =
\left \langle \left \langle \frac{\epsilon^g_{\bf j,l}}{\bar{n}_{\bf j,l}}
  \frac{\epsilon^g_{\bf j',l'} }{\bar{n}_{\bf j',l'}}
 \right \rangle_P \right \rangle
- \left \langle\frac{1}{\bar{n}_{\bf j,l}} \right \rangle -
\frac{L_1L_2L_3}{2^{j_1+j_2+j_3}}.
\end{equation}
The last term in eq.(37) is due to the normalization eq.(8).
 From eq.(37), the 2nd moment $D_{\bf j}$ is given by
\begin{equation}
D_{\bf j}=\frac{1}{L_1L_2L_3}
  \sum_{l_1=0}^{2^{j_1}-1} \sum_{l_2=0}^{2^{j_2}-1}\sum_{l_3=0}^{2^{j_3}-1}
  \left [ \left (\frac{\epsilon^g_{\bf j,l}}{\bar{n}_{\bf j,l}}\right )^2
    - \frac{1}{\bar{n}_{\bf j,l}} \right ] -1.
\end{equation}
The second term on the r.h.s. under the summation is the correction due to
Poisson noise, and the rest is the normalized
2nd moment of $\delta({\bf x})$. One can also calculate the
Poisson correction for higher order moments with eq.(33).

\section{Redshift distortion of one-point statistics}

\subsection{DWT CiC variables in redshift space}

Actually, astronomical data often gives only the distribution in redshift
space,
i.e.
\begin{equation}
n^S({\bf s})=
  \sum_{m=1}^{N_g}w_m\delta_D[{\bf s-x}_m- \hat{\bf r}v_r({\bf x}_m)/H]=
  \bar{n}^S({\bf s})[1+\delta^S({\bf s})],
\end{equation}
where $v_r({\bf x}_i)$ is the radial ($\hat{\bf r}$) component of
the velocity of the $i$th galaxy, $\bar{n}^S({\bf s})$ is the selection
function in redshift space, and $H$ is the Hubble constant at the
corresponding redshift.
Thus, we can only calculate the CiC moments
in redshift space.

For a given mass field $\delta({\bf x})$, the galaxy velocity
${\bf v}({\bf x})$ is a random field with mean
\begin{equation}
{\bf V}({\bf x})=\langle {\bf v}({\bf x})\rangle_v,
\end{equation}
where $\langle..\rangle_v$ is the average over the ensemble of
velocities. The mean velocity ${\bf V}({\bf x})$ is also called bulk
velocity at ${\bf x}$, which is assumed
irrotational (Bertschinger \& Dekel 1989; Dekel, Bertschinger \&
Faber 1990; Dekel et al. 1999).
In linear regime, the  bulk velocity is related
to the density contrast by
\begin{equation}
\delta({\bf x})= -\frac{1}{H\beta}\nabla\cdot {\bf V}({\bf x}),
\end{equation}
where the parameter $\beta \simeq \Omega^{0.6}/b$ at present, i.e.
redshift $z= 0$, and $b$ is the linear bias parameter.
The {\it rms} deviation of velocity ${\bf v}({\bf x})$ from the
bulk velocity ${\bf V}({\bf x})$ is
\begin{equation}
\langle [v_i({\bf x})-V_i({\bf x})]^2\rangle_v
  =[\sigma^{v_i}({\bf x})]^2, \mbox{ $i=1,2,3$}.
\end{equation}

In order to express the scale dependence of the bulk velocity
$V_i({\bf x})$ and the variance $\sigma^{v}({\bf x})$, we can also
decompose the velocity field  ${\bf v}({\bf x})$ with a DWT, i.e.
\begin{equation}
\epsilon^{v_i}_{\bf j, l}=\int v_i({\bf x})\phi_{\bf j, l}
  ({\bf x})d{\bf x}.
\end{equation}
Obviously, the variance $\sigma^{v}({\bf x})$ on scale ${\bf j}$
is given by
\begin{equation}
\sigma^{v}_{\bf j} = \left [\sum_{i=1}^{3}\langle v_i^2 \rangle_v
- \frac {2^{j_1+j_2+j_3}}{L_1L_2L_3}
\sum_{i=1}^{3} \langle \epsilon^{v_i}_{\bf j, l} \rangle_v^2
\right ]^{1/2}.
\end{equation}
Although $\epsilon^{v_i}_{\bf j, l}$ is ${\bf l}$-dependent,
$\sigma^{v}_{\bf j}$ should be ${\bf l}$-independent if the random field
${\bf v}({\bf x})$ is ergodic. On large scales, say
$\geq 10$h$^{-1}$Mpc, the velocity field ${\bf v}({\bf x})$ is roughly
\emph{Gaussian}, and it can be described by its mean and variance,
which are generally scale-dependent.

\subsection{Redshift distortion without selection function effect}

Because of eq.(39), the directly measurable DWT variable is not given by
eqs. (26) or (28), but
\begin{eqnarray}
\epsilon^{S}_{\bf j,l} & = &
\int \delta^S({\bf s})\phi_{\bf j,l}({\bf s})d{\bf s}
= \int\left [n^S({\bf s})/\bar{n}^S({\bf s}) - 1 \right]
\phi_{\bf j,l}({\bf s})d{\bf s}  \\ \nonumber
 & = & \frac{1}{\bar{n}^S_{\bf j,l}} \sum_{m=1}^{N_g}w_m
\phi_{\bf j,l}({\bf x}_m + \hat{\bf r} v_r({\bf x}_m)/H)
- \sqrt{\frac{L_1L_2L_3}{2^{j_1+j_2+j_3}}}.
\end{eqnarray}
where $\bar{n}^S_{\bf j,l}$ is calculated in the same way as
eq.(31), but with $n^S({\bf s})$ to replace $n^g({\bf x})$.

In this section, we do not consider the effect of selection
function, i.e. $\bar{n}=const$. In the plane-parallel
approximation, i.e. $\hat{{\bf r}}$ is along the $\hat{x_3}$-direction,
eq.(45) becomes
\begin{equation} \label{eq:epss-taylor}
\epsilon^{S}_{{\bf j,l}} =\frac{1}{\bar{n}}
\int n^g({\bf x})
e^{v_3({\bf x})\frac{1}{H} \frac{\partial}{\partial x_3}}
 \phi_{\bf j,l}({\bf x}) d {\bf x}
- \sqrt{\frac{L_1L_2L_3}{2^{j_1+j_2+j_3}}}.
\end{equation}
If the velocity field is Gaussian, subjecting eq.(46) to an average
over the ensemble of velocities, we have
\begin{eqnarray}
\langle \epsilon^{S}_{{\bf j,l}}\rangle_v &  = &
\frac{1}{\bar{n}} \int n^g({\bf x}) \langle
  e^{v_3({\bf x})\frac{1}{H} \frac{\partial}{\partial x_3}} \rangle_v
  \phi_{\bf j,l}({\bf x}) d{\bf x}
- \sqrt{\frac{L_1L_2L_3}{2^{j_1+j_2+j_3}}}  \\ \nonumber
& = & \int [1+\delta({\bf x})]
  e^{V_3({\bf x})\frac{1}{H} \frac{\partial}{\partial x_3} +
  \frac{1}{2}(\sigma_{\bf j}^{v_3}/H)^2 \left (\frac{\partial}
  {\partial x_3} \right)^2} \phi_{\bf j,l}({\bf x}) d{\bf x}
- \sqrt{\frac{L_1L_2L_3}{2^{j_1+j_2+j_3}}}.
\end{eqnarray}
For clarity, the angle brackets $\langle \ldots\rangle_v$ are dropped
hereafter without causing any confusion.

If we consider only the linear effect of the bulk velocity, equation
(47) is approximately
\begin{equation}
 \epsilon^{S}_{\bf j,l} \simeq \int [1+\delta({\bf x})]
\left [1+V_3({\bf x})\frac{1}{H} \frac{\partial}{\partial x_3} \right ]
 e^{\frac{1}{2}(\sigma_{\bf j}^{v_3}/H)^2 \left (\frac{\partial}
  {\partial x_3} \right)^2} \phi_{\bf j,l}({\bf x}) d{\bf x}
- \sqrt{\frac{L_1L_2L_3}{2^{j_1+j_2+j_3}}}.
\end{equation}
Neglecting the terms of the order of $V_3({\bf x})\delta({\bf x})$, and
using the linear relation between $\delta({\bf x})$ and ${\bf V}({\bf x})$,
we have
\begin{eqnarray} \label{eq:1st-order-delta}
\epsilon^{S}_{\bf j,l} &\simeq& \int \left\{1+\delta({\bf x})
- \beta \left[\nabla^{-2} \frac{\partial}{\partial x_3}\delta({\bf x})
\right ] \frac{\partial }{\partial x_3} \right\}
e^{\frac{1}{2}(\sigma_{\bf j}^{v_3}/H)^2 \left (\frac{\partial}
  {\partial x_3} \right)^2} \phi_{\bf j,l}({\bf x}) d{\bf x}
- \sqrt{\frac{L_1L_2L_3}{2^{j_1+j_2+j_3}}} \\ \nonumber
&=& \int e^{\frac{1}{2}(\sigma_{\bf j}^{v_3}/H)^2 \left (\frac{\partial}
  {\partial x_3} \right)^2} \phi_{\bf j,l}({\bf x}) d{\bf x}
+\sum_{{\bf l}'} \epsilon_{{\bf j,l}'} \int
\phi_{{\bf j,l}'}({\bf x}) e^{\frac{1}{2}(\sigma_{\bf j}^{v_3}/H)^2
\left (\frac{\partial}
  {\partial x_3} \right)^2} \phi_{\bf j,l}({\bf x}) d{\bf x}\\ \nonumber
& & +\beta \sum_{\bf l'} \epsilon_{\bf j,l'} \int
 \phi_{\bf j,l'}({\bf x})
 e^{\frac{1}{2}(\sigma_{\bf j}^{v_3}/H)^2 \left (\frac{\partial}
{\partial x_3} \right)^2}\nabla^{-2} \frac{\partial^2}{\partial x_3^2}
\phi_{\bf j,l}({\bf x}) d{\bf x} -
\sqrt{\frac{L_1L_2L_3}{2^{j_1+j_2+j_3}}}.
\end{eqnarray}
In the last step, integration by parts for $\partial/\partial x_3$ and
the completeness relation equation (10) are
used. The summation of ${\bf l}$ in eq.(49) is over
$l_i=0...2^{j_i-1}$.

To simplify eq.(49), we define the following matrices
\begin{equation}
\gamma^{a,b}_{\bf j, l,l'}=\int \phi_{\bf j,l}({\bf x}) A^aB^b
   \phi_{\bf j,l'}({\bf x}) d{\bf x},
\end{equation}
where the differential operators $A$ and $B$ are
\begin{equation}
A= e^{\frac{1}{2}(\sigma_{\bf j}^{v_3}/H)^2 \left (\frac{\partial}
{\partial x_3} \right)^2},  \hspace{4mm}
B= \nabla^{-2} \frac{\partial^2}{\partial x_3^2}.
\end{equation}
Obviously $\gamma^{a,b}_{\bf j, l,l'}$ depends on
$\Delta {\bf l} = (|l_1 - l'_1|, |l_2-l'_2|, |l_3-l'_3|)$, not ${\bf l}$
or ${\bf l'}$ individually.
Appendix A provides the algorithms to calculate
$\gamma^{a,b}_{\bf j, l,l'}$.

Thus, eq.(49) becomes
\begin{equation}
 \epsilon^{S}_{\bf j,l}  =
  - \sqrt{\frac{L_1L_2L_3}{2^{j_1+j_2+j_3}}}
  + \sqrt{\frac{L_1L_2L_3}{2^{j_1+j_2+j_3}}}
    \sum_{\bf l'}\gamma^{1,0}_{\bf j, l,l'}
  + \sum_{\bf l'}\gamma^{1,0}_{\bf j, l,l'} \epsilon_{\bf j,l'}
  + \beta \sum_{\bf l'}\gamma^{1,1}_{\bf j, l,l'} \epsilon_{\bf j,l'}
\end{equation}

Using the so-called ``partition of unity'' (Fang \& Feng 2000),
one can show that
\begin{equation}
\sum_{\bf l'} \gamma^{1,0}_{\bf j, l,l'} \simeq 1.
\end{equation}
Therefore, eq.(52) gives
\begin{equation}
 \epsilon^{S}_{\bf j,l}  \simeq
   \sum_{\bf l'}(\gamma^{1,0}_{\bf j, l,l'}
  + \beta \gamma^{1,1}_{\bf j, l,l'}) \epsilon_{\bf j,l'}.
\end{equation}
This is the mapping of one-point variables between real and redshift
spaces.

\subsection{Redshift distortion of the 2nd moment}

The second moment of one-point statistics in redshift space is
\begin{equation}
\langle |\epsilon^{S}_{\bf j,l}|^2\rangle \simeq
\Big\langle
[\sum_{\bf l'}(\gamma^{1,0}_{\bf j, l,l'}
  + \beta \gamma^{1,1}_{\bf j, l,l'}) \epsilon_{\bf j,l'}]
[\sum_{\bf l''}(\gamma^{1,0}_{\bf j, l,l''}
  + \beta \gamma^{1,1}_{\bf j, l,l''}) \epsilon_{\bf j,l''}]
 \Big\rangle.
\end{equation}
One can show that, even for a weakly non-linear field,
both $\langle \epsilon_{\bf j,l} \epsilon_{\bf j,l'}\rangle$ and
$\gamma^{a,b}_{\bf j, l, l'}$ are symmetric and
quasi-diagonalized with respect to ${\bf l}$ and ${\bf l'}$, and
$\langle |\epsilon_{\bf j,l}|^2\rangle$
should be ${\bf l}$-independent. The wavelet counterpart, i.e.
$\langle \tilde{\epsilon}_{\bf j, l}\tilde{\epsilon}_{\bf j, l'}
\rangle\simeq \delta^K_{\bf l,l'}
\langle \tilde{\epsilon}^2_{\bf j, l}\rangle$,
has been shown by Pando, Feng \& Fang (2001). Thus, eq.(55) becomes
\begin{equation}
\langle |\epsilon^{S}_{\bf j,l}|^2\rangle \simeq \left (
  \sum_{\bf l'}\gamma^{1,0}_{\bf j, l,l'}\gamma^{1,0}_{\bf j, l,l'}
 +2\beta \sum_{\bf l'}\gamma^{1,0}_{\bf j, l,l'}\gamma^{1,1}_{\bf j, l,l'}
 +\beta^2\sum_{\bf l'}\gamma^{1,1}_{\bf j, l,l'}\gamma^{1,1}_{\bf j, l,l'}
  \right ) \langle |\epsilon_{\bf j,l}|^2\rangle.
\end{equation}

Using the completeness relation equation (10), we have
\begin{equation}
\sum_{\bf l''} \gamma^{a_1,b_1}_{\bf j, l,l''}
\gamma^{a_2,b_2}_{\bf j, l'',l'} =
  \gamma^{a _1+ a_2, b_1+b_2}_{\bf j, l,l'}.
\end{equation}
Thus, eq.(56) gives
\begin{equation}
\langle |\epsilon^{S}_{\bf j,l}|^2\rangle \simeq (\gamma^{2,0}_{\bf j,  0,0}
  +2\beta\gamma^{2,1}_{\bf j, 0,0}
 +\beta^2\gamma^{2,2}_{\bf j, 0,0}) \langle |\epsilon_{\bf j,l}|^2\rangle,
\end{equation}
where we have used $\gamma^{a,b}_{\bf j, l,l}= \gamma^{a,b}_{\bf j, 0,0}$.
Therefore, we have finally the mapping of $D_{\bf j}$ between real and
redshift spaces
\begin{equation}
D^S_{\bf j}\simeq(\gamma^{2,0}_{\bf j, 0,0}
  +2\beta\gamma^{2,1}_{\bf j, 0,0}
 +\beta^2\gamma^{2,2}_{\bf j, 0,0})D_{\bf j}.
\end{equation}

\subsection{Effect of selection functions}

In linear approximation, it is reasonable to estimate the effect of
redshift distortion of $n^S({\bf s})$ and $\bar{n}^S({\bf s})$
separately. In this case, one can still use eq.(54) as the mapping
from $\epsilon^{S}_{{\bf j,l}}$ to $\epsilon_{{\bf j,l}}$.
We only need to study the effect of the mapping between
$\bar{n}^S({\bf s})$ and $\bar{n}({\bf x})$, which is given by
\begin{equation}
 \langle \bar{n}^S({\bf s})\rangle_v  =
  \langle \bar{n}[{\bf x}+ \hat{\bf r}v_r({\bf x})/H] \rangle_v
  \simeq \bar{n}({\bf x}) + \frac{1}{H}V_r({\bf x})
    \hat{\bf r}\cdot \nabla n^g({\bf x}),
\end{equation}
where we have used $\langle v_r \rangle_v=V_r$. With the plane-parallel
approximation of selection function (\S 3.2), eq.(60)  becomes
\begin{equation}
 \bar{n}^S({\bf s})=\bar{n}(x_3) +
    \frac{1}{H}\frac{d \bar{n}(x_3)}{dx_3}V_3({\bf x}),
\end{equation}
where we have dropped $\langle...\rangle_v$ for $\bar{n}^S({\bf s})$.
>From equation (41), $V_3$ can be represented by $\delta({\bf x})$, so
we have
\begin{eqnarray}
 \bar{n}^S({\bf s}) & = & \bar{n}(x_3)\left [ 1-
    \beta\frac{d\ln \bar{n}(x_3)}{dx_3}
     \frac{\partial}{\partial x_3}\nabla^{-2} \delta({\bf x}) \right ]
     \\ \nonumber
 & \simeq &
 \bar{n}(x_3) \left \{ 1 -
    \beta\frac{d\ln \bar{n}(x_3)}{dx_3}
     \frac{\partial}{\partial x_3}
    \nabla^{-2}\left [\frac{n^g ({\bf x})}{\bar{n}(x_3)} - 1
\right ] \right \}.
\end{eqnarray}

Combining eqs.(54) and (62), we have finally
\begin{equation}
\frac{\epsilon_{\bf j,l}^{S}}{\bar{n}^S_{\bf j,l}} \simeq
 \sum_{\bf l'}[\gamma^{0,1}_{\bf l,l'} + \beta\gamma^{1,1}_{\bf l,l'}]
   \frac{ \epsilon_{\bf j,l'}}{\bar{n}_{\bf j,l'}}
  +
  \left . \beta
   \frac{d\ln \bar{n}(x_3)}{dx_3}\right|_{\bf j,l}
   \sum_{l_3} Q_{\bf j,l,l'}
   \frac{\epsilon_{\bf j,l'}}{\bar{n}_{ \bf j,l' }},
\end{equation}
where $d\ln \bar{n}^S(x_3)/dx_3|_{\bf j,l}$ stands for the mean value
of $d\ln \bar{n}^S(x_3)/dx_3$ in the cell $({\bf j,l})$, and
$\bar{n}_{\bf j,l}$ is given by eq.(31). In the last term of
eq.(63), $l_1=l'_1$ and $l_2 =l'_2$, and the summation runs only
over $l_3$. The coefficient
$Q_{\bf j,l, l'}$ is defined by
\begin{equation}
 Q_{\bf j,l, l'}=\int \phi_{{\bf j},l_1,l_2,l_3}({\bf x})
 \frac{\partial}{\partial x_3}\nabla^{-2}\phi_{{\bf j},l_1,l_2,l'_3}
 ({\bf x}) d{\bf x}.
\end{equation}
The calculation of $Q_{\bf j,l, l'}$ is given in Appendix A.

Because all ${\bf l}$-diagonal elements of $Q_{\bf j,l,l'}$ are zero
(Appendix A), and
$\gamma^{a,b}_{\bf j,l,l'}$ are quasi-${\bf l}$-diagonal, the first
and the second
terms on the r.h.s. of equation (63) are not correlated. We have then
\begin{eqnarray}
\left \langle \left (
\frac{\epsilon_{\bf j,l}^{S}}{\bar{n}^S_{\bf j,l}} \right )^2
   \right \rangle &  = &
 (\gamma^{0,2}_{\bf j, 0,0}
  +2\beta\gamma^{1,2}_{\bf j, 0,0}
 +\beta^2\gamma^{2,2}_{\bf j, 0,0})\left \langle \left (
\frac{\epsilon_{\bf j,l}}{\bar{n}_{\bf j,l}} \right )^2\right \rangle
     \\ \nonumber
 &       + &
  \left [
\left . \beta\frac{d\ln \bar{n}(x_3)}{dx_3}\right|_{\bf j,l}\right ]^2
    \sum_{\bf l'} Q^2_{\bf l, l'}
  \left \langle \left (
 \frac{\epsilon_{\bf j, l'}}{\bar{n}_{\bf j,l'}}
   \right )^2 \right \rangle.
\end{eqnarray}
For a uniform field,
$\langle |\epsilon_{\bf j,l}^{S}/\bar{n}^S_{\bf j,l}|^2\rangle$
and $\langle |\epsilon_{\bf j,l}/\bar{n}_{\bf j,l}|^2\rangle$
are ${\bf l}$-independent. Thus, equation (65) gives
\begin{equation}
D^S_{\bf j}= \left ( \gamma^{0,2}_{\bf j,0,0}
  +2\beta\gamma^{1,2}_{\bf j,0,0}
 +\beta^2\gamma^{2,2}_{\bf j,0,0}+
 \left [
\left .\beta\frac{d\ln \bar{n}(x_3)}{dx_3}\right|_{\bf j,l}\right ]^2
    \sum_{\bf l'} Q^2_{\bf 0, l'} \right ) D_{\bf j}.
\end{equation}

Using the inequality equation (A16), we can show that if
\begin{equation}
 \frac{d\ln \bar{n}(x_3)}{dx_3} < \frac{2^{j_3}}{(2\pi)^{3/2}L_3},
\end{equation}
we have
\begin{equation}
 \left [ \left . \beta \frac{d\ln \bar{n}(x_3)}{dx_3}\right|_{\bf
j,l}\right ]^2
 \sum_{l_3-l'_3} Q^2_{{\bf j}, l_3-l'_3} < \beta^2 S^2_{\bf j}.
\end{equation}
That is, the selection function term in equation (66) is even less
than the second order terms $\beta^2 S^2_{\bf j}$ if the
selection function is slowly varying with $x_3$.

\section{Testing the DWT mapping between real and redshift spaces}
  \label{sec:test}

\subsection{Simulation samples}

We use N-body simulation samples to test the DWT algorithms for
recovering the one-point statistics from redshift space to real space.
The model parameters are listed in Table 1. Notice that the
simulation boxes are relatively large to minimize the effect on the
bulk velocities due to the missing power at long wavelengths
(Tormen \& Bertschinger 1996).

We use a modified P$^3$M code (Jing \& Fang 1994) to evolve $128^3$
($256^3$ for LCDM2) cold dark matter (CDM) particles
in a periodic cube of length $L$ on each side. The
linear power spectrum is given by the fitting formula in Bardeen et al.
(1986). Zel'dovich approximation is applied to set up
the initial perturbation.
The particles evolve 600 (800 for LCDM2) integration steps from
$z_i=12$ down to $z=0$ for all the models.

\begin{table*}
 \begin{center}
 \centerline{Table 1}
 \bigskip
 \begin{tabular}{cccccccc}
 \hline\hline
   Model& L/h$^{-1}$Mpc & $\Omega$ & $\Lambda$ &
     $\Gamma$ & $\sigma_8$ & run & particle\\
   \hline
  LCDM1  & 800 & 0.3 & 0.7 & 0.225 & 0.95 & 6 & $128^3$ \\
  LCDM2  & 800 & 0.3 & 0.7 & 0.21  & 0.81 & 1 & $256^3$ \\
  OCDM   & 800 & 0.3 & 0.0 & 0.225 & 0.95 & 6 & $128^3$ \\
  SCDM   & 800 & 1.0 & 0.0 & 0.50  & 0.62 & 6 & $128^3$ \\
  TCDM   & 800 & 1.0 & 0.0 & 0.25  & 0.60 & 6 & $128^3$ \\
\hline
 \end{tabular}
 \caption{Models of N-body simulations.\label{tab1}}
 \end{center}
\end{table*}

In addition to $\beta$ at $z \simeq 0$, we can also test the
redshift distortion algorithms at high redshifts. In this case,
$\beta$ is a function of $z$, $\Omega$ and $\Lambda$ as
(Lahav et al. 1991)
\begin{equation} \label{eq:beta}
\beta(z, \Omega, \Lambda ) \simeq
  \left [\frac{\Omega(1+z)^3}{\Omega(1+z)^3+
(1-\Omega-\Lambda)(1+z)^2+\Lambda}\right]^{0.6}.
\end{equation}
The simulation code is modified to generate light-cone outputs
from $z=3$ to $0$ by a similar method in the
{\it Hubble Volume Simulations} (Evrard et al. 2001). Instead of
producing spherical light-cones, we apply the plane-parallel
approximation in which a plane (light-front) sweeps through the simulation
box at the speed of light. Let the LOS be along the $x_3$-axis,
and the position of the light plane at time $t$ be
$x^L_3(t)$. The position and velocity of a particle is recorded when
it crosses the plane, i.e. when the position $x_3(t)$ of the particles
satisfies
\begin{equation}
x_3(t) = x^L_3(t).
\end{equation}
Since the time step $\Delta t$ in the simulations is finite,
it is computationally impractical to use eq.(70) directly.
The position of the
plane in time interval from step $i$ to step $i+1$ is approximately
\begin{equation}
x^L_3(t_i + \alpha \Delta t) \simeq x^L_3(t_i) +
\alpha[ x^L_3(t_{i+1}) - x^L_3(t_i)],
\end{equation}
where $\alpha$ is from 0 to 1. On the other hand, for a particle
we have $x_3(t_i+\alpha \Delta t ) \simeq x_3(t_i) +
 \alpha \Delta t \ v_3(t_i)$, where $v_3(t_i)$ is the velocity of the
particle along the LOS. Thus, if we can find a solution of
$|\alpha|<1$ from the following equation
\begin{equation}
 x_3(t_i) +  \alpha \Delta t \ v_3(t_i)=x^L_3(t_i) +
\alpha[ x^L_3(t_{i+1}) - x^L_3(t_i)],
\end{equation}
we have $x_3(t_i+\alpha \Delta t ) \simeq x^L_3(t_i + \alpha \Delta t)$,
i.e. the particle crosses the plane at time $t_i+\alpha \Delta t$.
We then record its position ${\bf x}(t_i) + \alpha \Delta t \
{\bf v}(t_i)$ and velocity ${\bf v}(t_i)$. The accuracy is
satisfactory. For example, the difference between
$x_3(t_i + \alpha \Delta t)$ and
$x_3^L(t_i + \alpha \Delta t)$ is typically less than 50 $h^{-1}$kpc out
to redshift $z = 2.5$ in the LCDM1 model. In the actual simulations,
the integration variable is $a$, the cosmic expansion factor, instead of
$t$.

Since the size of the simulation box is less than the distance swept
by the
plane from redshift $z=3$ to 0, the motion of the plane is realized
by periodic extensions of the simulation box when the plane meets
the boundary. This treatment should not have significant effects
on our analyses, because the simulations already impose a periodic boundary
condition, and the largest scale analyzed for each simulation is
only a quarter of its box size. Once the real-space light-cone
output is obtained, the observed redshift of a particle is given by
$z_{obs} = z + (1+z)v_r/c$, where $z$ is the actual redshift, and $v_r$
is the peculiar velocity. To avoid negative $z_{obs}$ caused by
peculiar velocities at low redshifts, a lower cut is set to $z=0.005$.
This results in less than a hundred particles bearing a negative
$z_{obs}$ in each run, which are then removed from the samples.
Three light-cone outputs along the three orthogonal axes are produced
for each simulation, which effectively increases the number of
realizations by a factor of 3.

\subsection{Calculation of $\gamma^{0,1}_{\bf j,0,0}$,
$\gamma^{0,2}_{\bf j,0,0}$, and $\gamma^{2,0}_{\bf j, 0,0}$}

Before testing the theory of recovering $D_{\bf j}$ (eq.(59)),
we need to study the factors $\gamma^{0,1}_{\bf j,0,0}$,
$\gamma^{0,2}_{\bf j,0,0}$, and
$\gamma^{2,0}_{\bf j, 0,0}$, because the first two are the key
of the redshift distortion of bulk velocity, while the last one the
distortion of velocity dispersion. Below we drop $({\bf 0,0})$
in subscripts for simplicity.

By the definition of eq.(50), the factors $\gamma^{0,1}_{\bf j}$ and
$\gamma^{0,2}_{\bf j}$ are given by
\begin{eqnarray}
\gamma^{0,1}_{\bf j}&=&\int \phi_{\bf j,0}({\bf x})
  \nabla^{-2} \frac{\partial^2}{\partial x_3^2}
  \phi_{\bf j,0}({\bf x})d{\bf x}, \\ \nonumber
\gamma^{0,2}_{\bf j}&=&\int \phi_{\bf j,0}({\bf x})
  \nabla^{-4} \frac{\partial^4}{\partial x_3^4}
  \phi_{\bf j,0}({\bf x})d{\bf x}.
\end{eqnarray}
 We plot $\gamma^{0,1}_{\bf j}$ and $\gamma^{0,2}_{\bf j}$ in Fig. 2 for
the D4 wavelet. It should be pointed out that $\gamma^{0,b}_{\bf j}$
depends only on the \emph{shape}, not the scale ${\bf j}$ of the scaling
function. For instance,
 $\gamma^{0,1}_{2,2,3} = \gamma^{0,1}_{3,3,4} = \ldots =
\gamma^{0,1}_{5,5,6}$, because the 3-D cells of ${\bf j}=(2,2,3),
(3,3,4) \ldots, (5,5,6)$ have the same shape, i.e. all the ratios of
length~:~width~:~height of these cells are
2:2:1, even though they have different volumes (scales).
For cubic cells, i.e. $j_1 = j_2 = j_3 = j$, we have
$\gamma^{0,1}_{j,j,j}\simeq 1/3$, and $\gamma^{0,2}_{j,j,j}\simeq 1/5$.

The factor $\gamma^{2,0}_{\bf j}$ is given by
\begin{equation}
\gamma^{2,0}_{\bf j} =\int \phi_{\bf j,0}({\bf x})
  e^{(\sigma_{\bf j}^{v_3}/H)^2 \left (\frac{\partial}
{\partial x_3} \right)^2}
  \phi_{\bf j,0}({\bf x})d{\bf x}.
\end{equation}
It depends on both the scaling function and the velocity dispersion
$\sigma_{\bf j}^{v_3}$, and therefore it is model-dependent.

The velocity variance $\sigma_{\bf j}^{v_3}$ is scale-dependent.
It is shown in Fig. 3 for all the models, which
is similar to Fig. 6 in Yang et al. (2002). Actually,
$\sigma^{v_3}_{\bf j}$ is given by all the power of pairwise
velocity on scales less than the scale of ${\bf j}$, and therefore,
it is larger on larger scales.

To avoid crowding the figure, we only show $\gamma^{2,0}_{j, j, j_3}$
for LCDM1 in Fig. 4. Other modes, such as $\gamma^{2,0}_{j_1, j_2, j_3}$,
follow $\gamma^{2,0}_{j, j, j_3}$ closely. An interesting
feature seen in Fig. 4 is that $\gamma^{2,0}_{j,j,2}$ and
$\gamma^{2,0}_{j,j,3}$ are almost equal to 1 from $j=2$ to 6, which
corresponds to spatial scales from 200 to 12.5 h$^{-1}$ Mpc on the
celestial  sphere, and 200 to 100 h$^{-1}$ Mpc along the LOS.
Even for $\gamma^{2,0}_{j,j,4}$,
the dependence on $j$ is only mild. Moreover, this property is
model-independent, i.e. for other models, we also have the
$j$-independence of $\gamma^{2,0}_{j,j,2}$ and
$\gamma^{2,0}_{j,j,3}$.

The factor $\gamma^{2,0}_{j_1, j_2, j_3}$ is sensitive to
$\sigma^{v_3}_{\bf j}$ (as a variable) when the scale along the LOS is
small, but insensitive when the scale is large.
For example, when $\sigma^{v_3}_{\bf j}$ varies from 0 to 500 km~s$^{-1}$,
$\gamma^{2,0}_{j,j,3}$ drops from 1 to 0.97, i.e. the change is only 3\%,
while $\gamma^{2,0}_{j,j,6}$ from 1 to 0.46, i.e. the change is by a
factor of 2. If we try to use a scale-independent $\sigma^v$ to recover
the real space power spectrum, we should chose $\sigma^v$ to give a good
fitting on small scales, because the value of $\sigma^v$ does not
significantly affect the large scales.

In addition to $j_3$, $\gamma^{2,0}_{\bf j}$ depends
on $j_1$ and $j_2$ (collectively $j_\perp$) through
$\sigma_{\bf j}^{v_3}$, which is less than
500 km~s$^{-1}$ (Fig. 3). Therefore, when the LOS scale is above
100 h$^{-1}$ Mpc, $\gamma^{2,0}_{\bf j}$ is almost a constant with
respect to $j_\perp$. On
the other hand, when the scale is below 50 h$^{-1}$ Mpc,
$\gamma^{2,0}_{\bf j}$ shows a mild dependence on $j_\perp$
due to the scale dependence of $\sigma_{\bf j}^{v_3}$.

In a word, we benefit from the DWT representation to see
the different behavior of the two types of the redshift distortion. 
The former is sensitive to the shape of the mode,
not the scale, while the latter just the contrary.

\subsection{Recovery of $D_{\bf j}$}

We neglect the effect of selection function here for the simulation
samples, which is similar to that on the DWT power spectrum tested in
Yang et al. (2002). Since the matrix $\gamma^{a,b}_{\bf j, l,l'}$
is quasi-diagonalized with respect to $({\bf l,l'})$, we have
approximately $\gamma^{a,b}_{\bf j} \simeq
\gamma^{a,0}_{\bf j}\gamma^{0,b}_{\bf j}$.  Thus, eq.(59) yields
\begin{equation}
D_{\bf j}^S \simeq
\gamma^{2,0}_{\bf j}(1 + 2\beta\gamma^{0,1}_{\bf j} +
\beta^2\gamma^{0,2}_{\bf j}) D_{\bf j}.
\end{equation}
In eq.(75), the contributions of the operators $A$ and $B$ to the redshift
distortion are separated. Usually terms of $B$ ($\gamma^{0,1}_{\bf j}$ and
$\gamma^{0,2}_{\bf j}$) are called linear redshift distortion, and that of
$A$ ($\gamma^{2,0}_{\bf j}$) non-linear distortion. These terminologies 
may cause a confusion. Here ``linear'' means only that $B$ is from 
the linear term of $V_3$ in the approximation eq.(48), and ``non-linear'' 
means that $A$ is non-linear of the velocity dispersion $\sigma^{v_3}$. 
It does not imply that the terms of $B$ is enough for estimating the 
redshift distortion of a field that is in the linear regime. Since the velocity 
dispersion $\sigma^v$ is non-zero in the linear regime, the non-linear 
distortion term of $A$ may also play a non-negligible role even when the 
field is linear. In this paper, 
we follow the tradition to call the terms of $B$ and $A$ the linear and 
non-linear effects of redshift distortion, respectively. However, it should 
be kept in mind that these names do not reflect whether the field is linear 
or non-linear.  

For ${\bf j}$-diagonal modes, equation (75) reduces to
$D_{\bf j}^S \simeq
\gamma^{2,0}_{\bf j}(1 + 2/3\beta + 1/5\beta^2) D_{\bf j}$. The
factors within the parentheses are known as the linear redshift-to-real
space mapping for two-point correlation function (Kaiser 1987).

Figure 5 shows the recovery of ${\bf j}$-diagonal $D_{\bf j}$. It is
evident that the recovery equation (75) works very well on all scales and
redshifts considered, except for the SCDM model at $z=0.71$. Due to
greater $\beta$ parameters, the distortions of the TCDM and the SCDM
models are generally stronger than that of the rest.
For these two models, $D^S_{\bf j}$ is almost parallel
to $D_{\bf j}$, which indicates a very weak non-linear
redshift distortion on all scales. This is consistent with the fact that
the velocity dispersions
of the two models are small on all scales compared to the others.
Since the volume of a cubic cell with $12.5$h$^{-1}$Mpc
on each side is approximately the same as that of a sphere with a radius of
$8$h$^{-1}$Mpc, the values of $D_{6,6,6}$ are consistent
with $\sigma^2_8$ at corresponding redshifts for each model. The
differences,
nevertheless, are due to the difference in the window functions.

Similar to $D_{\bf j}$, one can generalize $\sigma_8$ to $\sigma_R$,
which is the {\it rms} density fluctuation in a sphere of radius $R$.
Since the behavior of $\sigma_R^2$ is the same as $D_{j,j,j}$, one can
recover the real space $\sigma_R$
from the redshift space $\sigma_R^S$ in a similar way as eq.(75), i.e.
$\sigma_R^S \simeq [\gamma^{2,0}_{j,j,j}(1 + 2/3\beta + 1/5\beta^2)]^{1/2}
\sigma_R$, where $j$ is chosen to match the volumes of the DWT cell and
the spherical top-hat window of radius $R$. Figure 6
demonstrates the recovery of $\sigma_R$ for all the models. It is a
coincidence that $\sigma^S_8\simeq\sigma_8$ for the low density models.

Figures 7 and 8 give the recovery of off-diagonal $D_{\bf j}$.
This recovery works well on scales above 50 h$^{-1}$ Mpc. On scales
less than 50 h$^{-1}$ Mpc, the error is noticeable but still small. This
is partially due to the approximation of the quasi-diagonality of the
covariance
$\langle \epsilon_{\bf j, l}
\epsilon_{\bf j, l'}\rangle \simeq \delta^K_{\bf l,l'}
\langle \epsilon^2_{\bf j, l}\rangle$ in equation (56).
Moreover, the separation between the factors involving operators
$A$ and $B$ in eq.(75) is not perfect as well.

The errors of recovering $D_{\bf j}$ can be more clearly seen via a
comparison with the recovery of the DWT power spectrum
from redshift space $P_{\bf j}^S$ to real space $P_{\bf j}$. It is
(Yang et al. 2002)
\begin{equation}
P_{\bf j}^S \simeq
\Gamma^{2,0}_{\bf j}(1 + 2\beta\Gamma^{0,1}_{\bf j} +
\beta^2\Gamma^{0,2}_{\bf j}) P_{\bf j},
\end{equation}
where $\Gamma^{a,b}_{\bf j}$ is defined by eq.(50) but with replacement
of $\phi_{\bf j}$ to $\psi_{\bf j}$. The factors $\Gamma^{0,1}_{\bf j}$ and
$\Gamma^{0,2}_{\bf j}$ also satisfy $\Gamma^{0,1}_{j,j,j} \simeq 1/3$ and
$\Gamma^{0,2}_{j,j,j} \simeq 1/5$. The non-linear term
$\Gamma^{2,0}_{\bf j}$ behaves the same as $\gamma^{2,0}_{\bf j}$, but
with a little stronger dependence on ${\bf j}$.

We find that the recovery of $D_{\bf j}$ by eq.(75) has an error
within 10\% in most cases, while the DWT power spectrum recovery eq.(76) is
accurate to 5\% or better. This is because that the quasi-diagonality
of the SFC's covariance  $\langle
\epsilon_{\bf j, l}\epsilon_{\bf j, l'}\rangle \simeq \delta^K_{\bf l,l'}
\langle \epsilon^2_{\bf j, l}\rangle$
is poorer than the quasi-diagonality of the WFC's covariance
$\langle
\tilde{\epsilon}_{\bf j, l}\tilde{\epsilon}_{\bf j, l'}\rangle
\simeq \delta^K_{\bf l,l'}\langle \tilde{\epsilon}^2_{\bf j, l}\rangle$.
As we know from the DWT analysis, the covariance $\langle
\tilde{\epsilon}_{\bf j, l}\tilde{\epsilon}_{\bf j, l'}\rangle$
is always quasi-diagonal for a Gaussian field, regardless the power
spectrum\footnote{This is the property employed for data compression
of the DWT analysis (Louis, Maass and Rieder 1997)}. On the other hand,
the quasi-diagonality of the covariance $\langle
\epsilon_{\bf j, l}\epsilon_{\bf j, l'}\rangle$ is not a generic mathematical
property, but only an approximation.

\subsection{Problems with real samples}

All the above algorithms can be applied to real samples as well. 
However, there are several problems to be considered in the 
analysis on real data.

{\it 1. Data assignment on grid}

To calculate the moments of counts-in-cells, one has to cast a grid on the 
data set either explicitly or implicitly, and assign the data in cells 
defined by the grid. The arbitrariness of the 
assignment may lead to significant error (Colombi, Bouchet \& Schaeffer 
1995). One way to reduce this error is to shift the grid.
It is well known that this assignment may also result in spurious 
features of the power spectrum on scales around the Nyquist frequency 
of the grid. In the DWT analysis, one can use different grids, but
no shift is needed. For a given grid, the assignment is realized 
by the same scaling function used for the data decomposition. Since scaling 
functions are orthogonal and complete in each given scale ${\bf j}$, the 
spurious features and false correlations can be completely avoided (Fang 
\& Feng 2000). To test the effect of the assignment on the one point 
statistics, we calculate $D_{j,j,j}$ for a snapshot sample of LCDM1 with 
randomly shifted grids. The snapshot sample is take at $z = 0.11$. In Fig. 
9, Snap-0 represents $D_{j,j,j}$ without grid shifting, while Snap-32 and 
Snap-1024, respectively, the average over 32, and 1024 
random shifts for grids on all scales. No significant deviation is detected 
among the results. 

{\it 2. Boundary condition and edge effect}

In a real survey, the edge effect is unavoidable due to the geometry (e.g. 
Szapudi \& Colombi 1996). Since the DWT bases are localized in physical 
space, the effect of edges can be effectively suppressed by dropping modes 
that are close to the boundary of the samples. In other words, when 
calculating $D_{\bf j}$, the averaging in eq.(17) runs only over modes 
that are not significantly affected by the edge effect,
and the normalization factor $L_1L_2L_3$ is replaced by the volume over
which the average takes place. This method has been tested 
numerically in the power spectrum detection (Pando \& Fang 1998b). It shows 
that the power spectrum can be fairly reconstructed regardless whether 
applying a 
periodic boundary or zero padding outside of the samples. The treatment of 
dropping edge modes has also been successfully employed to obtain the power 
spectrum for the Las Campanas redshift survey of galaxies, which has a 
slice-like geometry (Yang et al. 2001b). When the interested scale is close
to the dimension of the sample, one cannot afford to drop all the edge modes, 
and then a more thorough treatment is needed.
The edge effect exists in our 
simulation sample because the light-cone output is not periodic along the
redshift axis, while we have used periodic D4 wavelets in the analysis.
For comparison, the real-space light-cone $D_{j,j,j}$ from Fig. 5 is included 
in Fig. 9. There is no significant difference between the results from 
the snapshot, which is truly periodic, and that from the light-cone around
the same redshift. The slightly larger standard deviation of the light-cone
result at 200 $h^{-1}$Mpc (a quarter of the simulation box) is due to a 
larger fraction of edge cells.

{\it 3. Non-Poisson sampling}

In \S 3.3, we have considered the correction for Poisson sampling. It is 
sufficient for simulation samples. However, it may not be typical for real 
samples. For instance, some galaxy catalogs may be given by sub-Poisson 
sampling on small scales, or small halos (e.g. Bullock, Wechsler \& 
Somerville, 2002). The sub-Poisson distribution is simply due to very low
mass of the considered halo, so that it cannot host any additional 
objects. In other words, the sub-Poisson distribution is significant on
small scales on which galaxies are anticorrelated. Therefore, this sub-Poisson 
sampling is similar to the sub-Poisson distribution of the 
Fermi-Dirac statistics (one state can host no more than one particle). 
The effect of this sub-Poisson has been extensively studied in quantum optics 
(e.g. Martin \& Landauer, 1992). It is possible to extend the Poisson 
sampling eq.(35) to include Fermi-like sampling. The algorithm for a modified 
Poisson sampling has been developed by Jamkhedkar, Bi and Fang (2001) 
(see their Append B). 

\section{$\beta$ estimators with the DWT moment of one-point statistics}
\label{sec:beta}

\subsection{$\beta$ estimator with scale-decomposed quantities}

We first consider a $\beta$ estimator without non-linear effect.
For instance, $\gamma^{2,0}_{j, j, 2}\simeq \gamma^{2,0}_{j,j,3} \simeq 1$
(Fig. 4), i.e. the non-linear effects are small above 100 h$^{-1}$
Mpc along the LOS, eq.(75) gives
\begin{equation}
D_{j_1, j_2, j_3}^S \simeq (1 + 2\beta\gamma^{0,1}_{j_1, j_2, j_3}+
\beta^2\gamma^{0,2}_{j_1, j_2, j_3}) D_{j_1, j_2, j_3},
\quad j_3 = 2,3.
\end{equation}
It is easy to construct a $\beta$ estimator with eq.(77), because
the quantities $D_{\bf j}$, $D^S_{\bf j}$, and $\gamma^{a,b}_{\bf j}$
are not rotationally invariant, but they satisfy the following symmetries
with respect to the triple indices $(j_1,j_2,j_3)$.
\begin{enumerate}
\item If the cosmic density and velocity fields are statistically
isotropic, $D_{\bf j}$ in real space is invariant with
respect to cyclic permutations of index  ${\bf j}=(j_1,j_2,j_3)$,
i.e.
\begin{equation}
 D_{j_1,j_2,j_3}=D_{j_3,j_1,j_2}=D_{j_2,j_3,j_1}.
\end{equation}
\item In the plane-parallel approximation, i.e. the coordinate $x_3$ is in
the redshift direction, we have
\begin{equation} \label{eq:D-cyc}
 D^S_{j_1,j_2,j_3}=D^S_{j_2,j_1,j_3},
\end{equation}
\begin{equation} \label{eq:gamma-cyc}
 \gamma^{a,b}_{j_1,j_2,j_3}=\gamma^{a,b}_{j_2,j_1,j_3}.
\end{equation}
\end{enumerate}
Using eqs.(77)-(80), we have a $\beta$ estimator as follows
\begin{equation}
\frac {D^S_{j,2,3}}{D^S_{j,3,2}}\simeq
\frac{1+ 2\beta\gamma^{0,1}_{j, 2, 3}+
\beta^2\gamma^{0,2}_{j, 2, 3}}{1
+ 2\beta\gamma^{0,1}_{j, 3, 2}+
\beta^2\gamma^{0,2}_{j, 3, 2}}.
\end{equation}
Eq.(81) looks very similar to the $\beta$ estimator with
quadrupole-to-monopole ratio, or the multipole moments of two-point
correlation function. However, eq.(81)
contains not only the information of shape (like multipole moments), but
also
scales. All the DWT quantities in eq.(77) are on scale ${\bf j}$.
That is, the $\beta$ in eq.(77) refers to mode ${\bf j}=(j_1,j_2,j_3)$.
Therefore, if $\beta$ is scale-dependent, the $\beta$ estimated by eq.(81)
is its value referring to ${\bf j}$. On the other hand, if $\beta$ is
scale-free, the values of $\beta$ given by estimator eq.(81) with different
mode ${\bf j}$ should be the same.

Figure 10 plots the results in the case of a scale-free $\beta$ for the LCDM1
model using estimators with $j=2,3,4$. The results indeed show that the
estimated
$\beta$ is independent of the ${\bf j}$
used in the estimator, and it follows the theoretical curves in the entire
redshift range considered. Therefore, we expect that the estimator
(81) would be useful to study the scale-dependence of bias
parameters of galaxies.

For comparison, fig 11 shows the $\beta$ estimated from
quadrupole-to-monopole ratio as
\begin{equation}
\frac{P_2^S(k)}{P_0^S(k)} = \frac{\frac{4}{3}\beta + \frac{4}{7}\beta^2}
{1+\frac{2}{3}\beta+\frac{1}{5}\beta^2},
\end{equation}
where $P_2^S(k)$ and $P_0^S(k)$ are the quadrupole moment and the
monopole moment of the power spectrum  $P^S({\bf k})$ respectively
(Cole, Fisher \& Weinberg 1994, hereafter CFW; Hamilton 1997).
We average the results from wavelength 50 h$^{-1}$Mpc to 400 h$^{-1}$Mpc
for each realization. Thus the error bars are among different realizations
of each model. As noticed in CFW, this
estimator gives a lower $\beta$ than the true value at
$z \simeq 0$ even for wavelengths $> 50$ h$^{-1}$Mpc. In addition,
we find that it progressively overestimates $\beta$ at higher and
higher redshifts. It is suggested in CFW that the underestimate at short
wavelengths is due to the impact of non-linear gravitational clustering.
However, this explanation is still difficult to apply for the underestimate at 
long wavelengths, and the overestimate at high redshift, as cosmic field on
long wavelengths or high redshifts should be linear.   

This discrepancy occurs because the quadrupole-to-monopole ratio 
eq.(82) is scale-dependent even when the field is in linear regime.
This scale-dependence is given by the non-linear redshift distortion,
or the distortion of velocity dispersion. As mentioned in \S 5.3, 
$\sigma^{v_3}$ is non-zero in linear clustering regime, the non-linear 
redshift distortion (or the distortion of $\sigma^{v_3}$) should be 
considered regardless whether the field is linear or non-linear. Therefore, 
the discrepancy in CFW cannot be eliminated for long 
wavelengths or high redshift. To solve this problem,  extra 
information, such as the power-spectrum index of the density fluctuations, 
or the real-space correlation function, is added in the 
quadrupole-to-monopole ratio estimator eq.(82) (Peacock et al. 2001).

\subsection{$\beta$ estimator considering with non-linear redshift distortion}

Different from the quadrupole-to-monopole ratio, the DWT $\beta$ 
estimators are scale-decomposed. It is useful to consider scale-dependent 
effect. The DWT estimators are able to consider the non-linear redshift 
distortion ($\sigma^{v_3}$ redshift distortion) without assuming extra 
information. Actually, the result with estimator eq.(81) has already shown 
the effect of the non-linear redshift distortion.
We can see from Fig. 10 that the $\beta$ estimated by eq.(81) are
slightly dependent on $j$, and it is progressively higher in the
order $j = 2,3,4$. This can be explained with Fig. 4, which shows that
the factor $\gamma^{2,0}_{\bf j}$ is less than one, and is smaller for 
a smaller scale. Thus the
estimator eq.(81), which ignores the factor $\gamma^{2,0}_{\bf j}$ in
eq.(77), leads to a progressively higher $\beta$ in the order $j = 2,3,4$.
Thus, a simplest way to estimate the non-linear effect
is to take an average over $\beta$ given by eq.(81) with different
$j$. The left panel of Figure 12 presents the averaged $\beta$ from Fig. 10
for the four models.
The error bars contain the contribution of the non-linear effect.
That is, the DWT algorithm can
do a self-test on whether the result is largely
affected by the non-linear effect.

More delicate $\beta$ estimators can be constructed if we consider
the following properties: the non-linear factor $\gamma^{2,0}_{\bf j}$
depends only on $j_3$, but independent of $j_\perp$ if the LOS
scale is above 100 h$^{-1}$Mpc, while
the linear factors $\gamma^{0,1}_{\bf j}$ and $\gamma^{0,2}_{\bf j}$
depends only on the shape of the DWT mode (\S 5.2).
These properties apply to $\Gamma^{2,0}_{\bf j}$, $\Gamma^{0,1}_{\bf j}$,
and $\Gamma^{0,2}_{\bf j}$ as well.
Thus, we can combine modes with similar $j_3$ but different shapes to
cancel the non-linear factors, such as
\begin{equation}
 \frac{\Gamma^{2,0}_{j,2,3}\Gamma^{2,0}_{j',3,2}}
      {\Gamma^{2,0}_{j',2,3}\Gamma^{2,0}_{j,3,2}} \simeq 1,
\quad j \neq j'.
\end{equation}
Thus, from eq.(76), we have a $\beta$
estimator as
\begin{equation}
 \frac{P^S_{j,2,3}P^S_{j',3,2}}{P^S_{j',2,3}P^S_{j,3,2}}
   =\frac{(1+2\beta \Gamma^{0,1}_{j,2,3}+\beta^2\Gamma^{0,2}_{j,2,3})
(1+2\beta \Gamma^{0,1}_{j',3,2}+\beta^2 \Gamma^{0,2}_{j',3,2})}
      {(1+2\beta \Gamma^{0,1}_{j',2,3}+\beta^2 \Gamma^{0,2}_{j',2,3})
(1+2\beta \Gamma^{0,1}_{j,3,2}+\beta^2 \Gamma^{0,2}_{j,3,2})}.
\end{equation}
This estimator is similar to that in Yang et al. (2002). It can be used
for any pairs ($j$ $\neq$ $j'$) even when the scale of $j$ is small.
The right panel of Figure 12 shows that for the
four models, the $\beta$ estimated by eq.(84) has no more than about
15\% error at all redshifts $z < 3$. One can also use the modes $(j, 3,4)$,
$(j,4,3)$ to replace modes $(j,2,3)$ and $(j,3,2)$ in eqs.(83) and (84), 
which gives similar results to Fig. 12.
That is, the weakly non-linear redshift distortion can be considered
by the DWT $\beta$
estimators on scales until about $50$ h$^{-1}$ Mpc when the
quadrupole-to-monopole estimator shows significant errors (Fig. 11).

The counterpart of eq.(84) with $D_{\bf j}$,
\begin{equation}
 \frac{D^S_{j,2,3}D^S_{j',3,2}}{D^S_{j',2,3}D^S_{j,3,2}}
   =\frac{(1+2\beta \gamma^{0,1}_{j,2,3}+\beta^2\gamma^{0,2}_{j,2,3})
(1+2\beta \gamma^{0,1}_{j',3,2}+\beta^2 \gamma^{0,2}_{j',3,2})}
      {(1+2\beta \gamma^{0,1}_{j',2,3}+\beta^2 \gamma^{0,2}_{j',2,3})
(1+2\beta \gamma^{0,1}_{j,3,2}+\beta^2 \gamma^{0,2}_{j,3,2})},
\end{equation}
is not as good as eq.(84). It causes an error of about 40\% in $\beta$
because again the quasi-diagonality of the SFC's covariance is poorer than
that of WFC's. Since a DWT decomposition
of a random field yields variables for calculating both the power spectrum
and one-point statistics, the $\beta$ estimations with eqs.(81) and (84)
can be done in the same time.

An accurate $\beta$ estimator requires the knowledge of a precise recovery
of $P_{\bf j}$ or $D_{\bf j}$, which accounts for the non-linear
redshift distortion due to the velocity dispersion, and even the second
order
effect of the bulk velocity ignored in eq.(48). In other words, the
non-linear redshift distortion exists even on scales that are linear in the
sense of structure formation, and it must be corrected to achieve a reliable
and accurate $\beta$.

\section{Discussion and Conclusion}

We have developed the one-point statistics of a perturbed density
field with the multiresolutional decomposition based on discrete wavelet
transform. Since the scale and shape of the DWT bases are well
defined, this frame work is very effective to deal with problems of how the
one-point distribution and its moments depend on the scale and shape of the
window function. With this property, we have established the algorithm of
one-point variable and its moments in considering the effects of Poisson
sampling and selection function. We have also established the algorithm for
recovering the DWT one-point statistics from the redshift distortion due to
bulk velocity, velocity dispersion and selection function. 

Because the recovery of the real-space DWT one-point variable
and its moments can be realized scale-by-scale, one can design $\beta$
estimators which are sensitive to the scale-dependence of $\beta$,
for instance, caused by the scale-dependence of bias parameter of
galaxies. These
$\beta$ estimators are effective in avoiding the difficulty caused by
the scale-dependence of the non-linear redshift distortion. Compared with
conventional $\beta$ estimators (Peacock et al 2001), the DWT
$\beta$ estimators do not need to assume that the velocity dispersion is
scale-independent, or to add extra information, such as the power-spectrum
index or the real-space correlation function of the field. Numerical
tests by N-body simulation samples show that the proposed estimators can
yield the correct value of $\beta$ with about 15\% uncertainty for all
popular CDM models in the redshift range $z\leq 3$.

Since DWT decomposition contains two sets of bases, the scaling function
$\phi_{j,l}(x)$ and wavelet $\psi_{j,l}(x)$, a DWT decomposition of a
density field $\delta(x)$ actually yields variables for one-point statistics
$\epsilon_{j,l}=\int \phi_{j,l}(x)\delta(x)dx$ as well as the variables for
calculating the power spectrum
$\tilde{\epsilon}_{j,l}=\int \psi_{j,l}(x)\delta(x)dx$ (Fang \& Feng 2000).
In this sense, one can say that the DWT decomposition unifies the
algorithm of CiC and power spectrum, which are only two aspects of the
statistics with the DWT variables SFCs and WFCs respectively. For a finite
size sample, these two aspects of statistics can play different roles: the
former contains information of the perturbations on scales larger than
the size, while the latter does not. However, the latter generally has
diagonalized covariance, while the former does not. Therefore, the
$\beta$ estimators with WFCs are better than SFCs, while SFCs are
useful to estimate the effect of perturbations on large scales.

Actually, the DWT decomposition can
provide more types of statistics variables. For 2-D or 3-D samples,
we have variables defined as
\begin{equation}
\int \delta({\bf x})
\phi_{j_1,l_1}(x_1)\psi_{j_2,l_2}(x_2)\psi_{j_3,l_3}(x_3)d{\bf x},
\end{equation}
\begin{equation}
\int \delta({\bf x})
\phi_{j_1,l_1}(x_1)\phi_{j_2,l_2}(x_2)\psi_{j_3,l_3}(x_3)d{\bf x}.
\end{equation}
Obviously, these variables are sampled partially by the scaling function
$\phi_{j,l}(x)$, and partially by the wavelet $\psi_{j,l}(x)$. Statistics
with these variables are not typical CiC, or power spectrum. They are,
however, useful to study the one-point statistics, or power spectrum
of 2-D and 3-D samples. With the method developed in this paper, it is
not difficult to calculate various corrections (Poisson sampling,
selection function, redshift distortion) on the one-point statistics with
variables (86) or (87).

\acknowledgments

HZ is grateful to Daniel Eisenstein for extensive discussions on
this paper and facilitating the LCDM2 simulation. HZ would also like to
thank David Burstein for hosting the utility codes in this paper.

\appendix

\section{Calculations of $\gamma^{a,b}_{\bf j}$, and
$Q_{\bf j,l,l'}$}

\subsection{$\gamma^{0,b}_{\bf j}$}

Let us consider the plane-parallel approximation, i.e. coordinate
$x_3$ is in the redshift direction. By definition eq.(50),
we have
\begin{equation}
\gamma^{0,b}_{\bf j} = \int \phi_{\bf j,l}({\bf x})
\frac{\partial^{2b}}{\partial x_3^{2b}}\nabla^{-2b}\phi_{\bf j,l}({\bf
x}) d{\bf x}.
\end{equation}
Because 1-D scaling function $\phi_{j,l}(x)$ is given by dilating and
translating the basic wavelet $\phi(\eta)$ as
\begin{equation}
 \phi_{j,l}(x)=\left (\frac{2^j}{L}\right )^{1/2} \phi(\frac{2^jx}{L}-l),
\end{equation}
the Fourier transform of $\psi_{j,l}(x)$ is
\begin{equation}
 \phi_{j,l}(x)=\frac{1}{L}\sum_{n=-\infty}^{\infty}
 \hat{\phi}_{j,l}(n)e^{-i2\pi n x/L},
\end{equation}
and
\begin{equation}
 \hat{\phi}_{j,l}(n)=\left (\frac{L}{2^j}\right )^{1/2}
     \hat{\phi}(n/2^j)e^{-i2\pi n l/2^j},
\end{equation}
where $\hat{\phi}(n)$ is the Fourier transform of the basic
scaling function
\begin{equation}
 \hat{\phi}(n)=\int_{-\infty}^{\infty} \phi(\eta)e^{-i2\pi n\eta}d\eta.
\end{equation}
The function $|\hat{\phi}_{j,l}(n)|^2$ is shown in Fig. 1.

Thus, equation (A1) becomes
\begin{equation}
 \gamma^{0,b}_{j_1,j_2,j_3}= \frac{1}{2^{j_1+j_2+j_3}}
   \sum_{n_1,n_2,n_3 = \infty}^{\infty}
   \left (\frac{k_3}{k}\right )^{2b}
   |\hat{\phi}(u_1)\hat{\phi}(u_2)\hat{\phi}(u_3)|^2,
\end{equation}
where vector ${\bf k}=2\pi(n_1/L_1, n_2/L_2, n_3/L_3)$, and
${\bf u} = (n_1/2^{j_1},n_2/2^{j_2},n_3/2^{j_3})$.
Since $\hat{\phi}(n)$ is non-zero only around $|n| \lesssim 1$, the
summation of equation (A6) actually only over
numbers of $|n_i|\lesssim 2^{j_i}$.

If $L_1=L_2=L_3=L$, equation (A6) becomes
\begin{equation}
  \gamma^{0,b}_{j_1,j_2,j_3}= \frac{1}{2^{j_1+j_2+j_3}}
    \sum_{n_1,n_2,n_3 = \infty}^{\infty}
   \left (\frac{n_3}{n}\right )^{2b}
   |\hat{\phi}(u_1)\hat{\phi}(u_2)\hat{\phi}(u_3)|^2.
\end{equation}

\subsection{$\gamma^{2,0}_{\bf j}$}

If $\sigma^v({\bf x})$ is independent of ${\bf x}$, $\gamma^{2,0}_{\bf j}$
is
\begin{equation}
 \gamma^{2,0}_{j_1,j_2,j_3} = \frac{1}{2^{j_1+j_2+j_3}}
    \sum_{n_1,n_2,n_3 = -\infty}^{\infty}
   |\hat{\phi}(u_1)\hat{\phi}(u_2)\hat{\phi}(u_3)|^2
     \exp[-(\sigma^{v_3}/H)^2(\hat{r}\cdot {\bf k})^2].
\end{equation}
In the plane-parallel approximation, we have
\begin{equation}
\gamma^{2,0}_{j_1,j_2,j_3} = \frac{1}{2^{j_3}}
   \sum_{n_3 = -\infty}^{\infty}
 |\hat{\phi}(u_3)|^2 \exp[-(\sigma^{v_3} k_3/H)^2].
\end{equation}
The summation of equations (A8) and (A9) also runs only over
numbers of $|n_i| \lesssim 2^{j_i}$.

\subsection{$\gamma^{a,b}_{\bf j, l, l'}$}

Using the results for $\gamma^{0,b}_{\bf j}$ and $\gamma^{2,0}_{\bf j}$,
it is easy to find
\begin{eqnarray}
\gamma^{a,b}_{\bf j,l,l'} & = & \frac{1}{2^{j_1+j_2+j_3}}
  \sum_{n_1,n_2,n_3 = -\infty}^{\infty}
\left (\frac{k_3}{k}\right )^{2b} \\ \nonumber
 & &
   \cos [2\pi {\bf u} \cdot ({\bf l}'-{\bf l})]
   \exp[-(a/2)(\sigma^{v_3} k_3/H)^2]
  |\hat{\phi}(u_1)\hat{\phi}(u_2)\hat{\phi}(u_3)|^2.
\end{eqnarray}
It is obvious from equation (A10)
that $\gamma^{a,b}_{\bf j, l, l'}$ is symmetric with respect to
${\bf l}$ and ${\bf l'}$, and it depends only on the difference
$(|l_1 - l'_1|, |l_2-l'_2|, |l_3-l'_3|)$. Consequently,
$\gamma^{a,b}_{\bf j, l, l}$ is independent of ${\bf l}$. In addition,
the elements $\gamma^{a,b}_{\bf j, l, l}$
are dominant over $\gamma^{a,b}_{\bf j, l, l'}$ with
${\bf l} \neq {\bf l'}$,
because the latter sums over oscillating terms.

\subsection{$Q_{\bf j,l,l'}$}

The quantity $Q_{\bf j,l,l'}$ is given by eq.(64)
\begin{equation}
 Q_{\bf j, l, l'}=\int \phi_{{\bf j},l_1,l_2,l_3}({\bf x})
 \frac{\partial}{\partial x_3}\nabla^{-2}
  \phi_{{\bf j},l_1,l_2,l'_3}({\bf x}) d{\bf x}.
\end{equation}
For index ${\bf l,l'}$, It depends only on the difference
$l_3-l'_3$. We have
\begin{equation}
 Q_{\bf j, l,l'} = Q_{{\bf j}, l_3-l'_3}=\frac{1}{2^{j_1+j_2+j_3}}
 \sum_{n_1,n_2,n_3 = \infty}^{\infty}  \frac{k_3}{k^2}
 \sin[2\pi u_3(l_3-l_3')]
|\hat{\phi}(u_1)\hat{\phi}(u_2)\hat{\phi}(u_3)|^2.
\end{equation}
Equation (A12) gives
\begin{equation}
  Q_{{\bf j}, l_3-l'_3}=0, \ \ \ {\rm if} \ \  l_3-l'_3=0.
\end{equation}

Since
\begin{equation}
\sum_{l_3-l'_3=0}^{2^{j_3}-1}
\sin[2\pi u_3(l_3-l_3')]\sin[2\pi u'_3(l_3-l_3')]
\simeq \left \{ \begin{array}{ll}
                1/2\pi u_3 & \mbox{if  $u_3=u'_3$} \\
                 0          & \mbox{otherwise}
               \end{array}
         \right .,
\end{equation}
we have
\begin{equation}
\sum_{l_3-l'_3}Q^2_{{\bf j}, l_3-l'_3}  =
  \frac{1}{2^{2(j_1+j_2+j_3)} }
  \frac {1}{2\pi u_3}\sum_{n_1,n_2,n_3=-\infty}^{\infty}
 \left (\frac{k_3} {k^2}\right)^2
|\hat{\phi}(u_1)\hat{\phi}(u_2)\hat{\phi}(u_3)|^4.
\end{equation}
Thus, equations(A7) and (A15) yield
\begin{equation}
 \sum_{l_3-l'_3}Q^2_{{\bf j}, l_3-l'_3} <
  \left [\frac{(2\pi)^{3/2}L_3}{2^{j_3}}\right ]^2 [\gamma^{0,1}_{\bf j}]^2.
\end{equation}

\newpage

\begin{figure}
\plotone{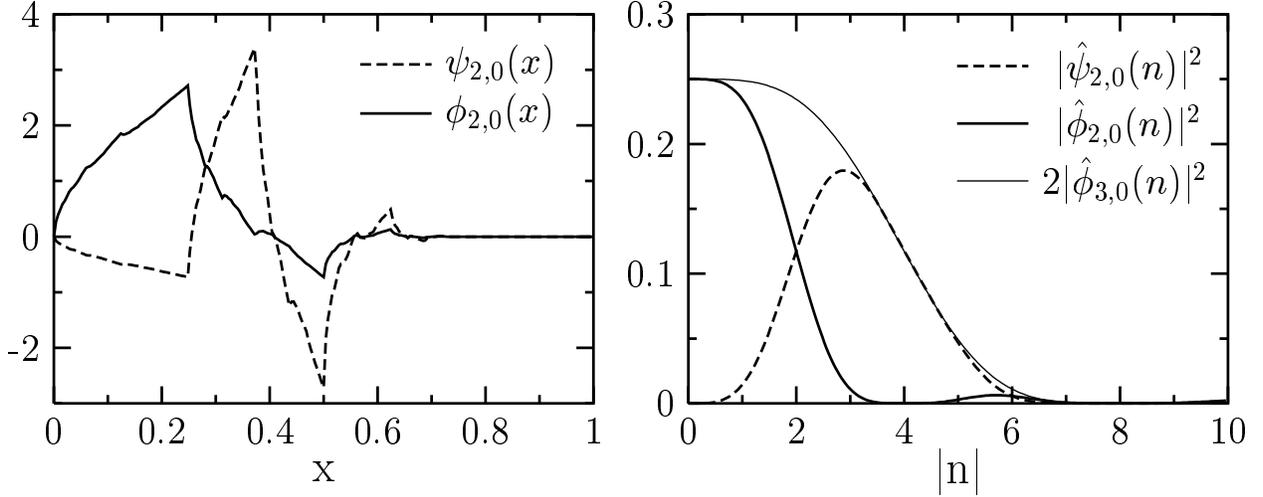}
\caption[f1.eps] {The scaling function $\phi_{2,0}(x)$, the
wavelet $\psi_{2,0}(x)$, and their Fourier transforms
$|\hat{\phi}_{2,0}(n)|^2$, $|\hat{\phi}_{3,0}(n)|^2$, and
$|\hat{\psi}_{2,0}(n)|^2$, where $n = k L/2\pi$, and $L=1$.
\label{Fig1} }
\end{figure}

\begin{figure}
\plotone{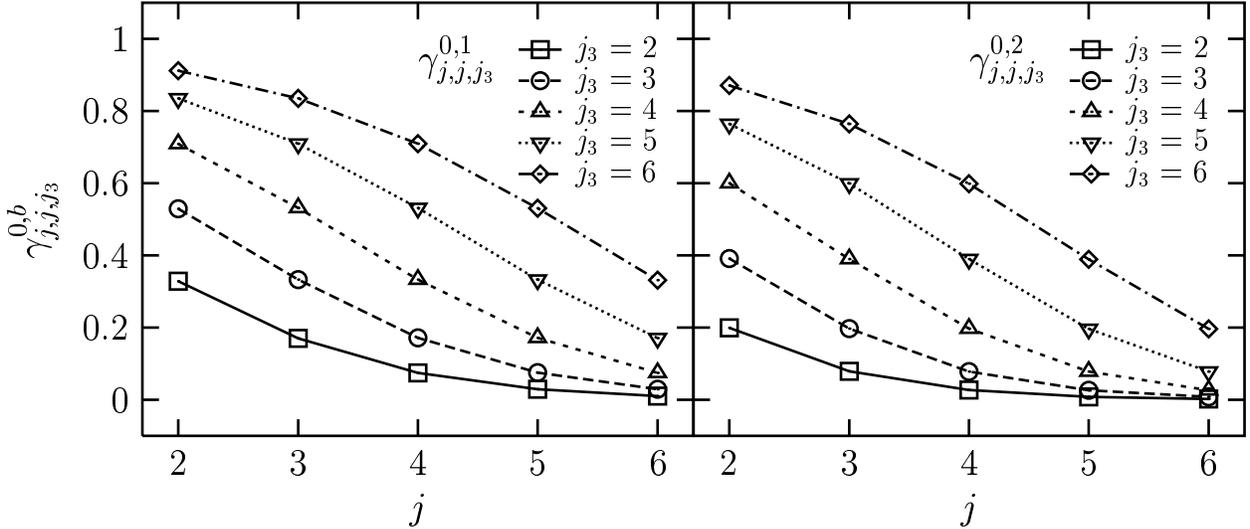}
\caption[f2.eps] {The factors of linear redshift distortion
$\gamma^{0,1}_{\bf j}$ and $\gamma^{0,2}_{\bf j}$ in equation
(73). The subscript $j_3$ indicates the $\hat{x}_3$-direction,
or the redshift direction. This convention is followed
in all the figures. These factors depend only on the geometry
of the DWT cells. }
\label{Fig2}
\end{figure}

\begin{figure}
\epsscale{0.9}
\plotone{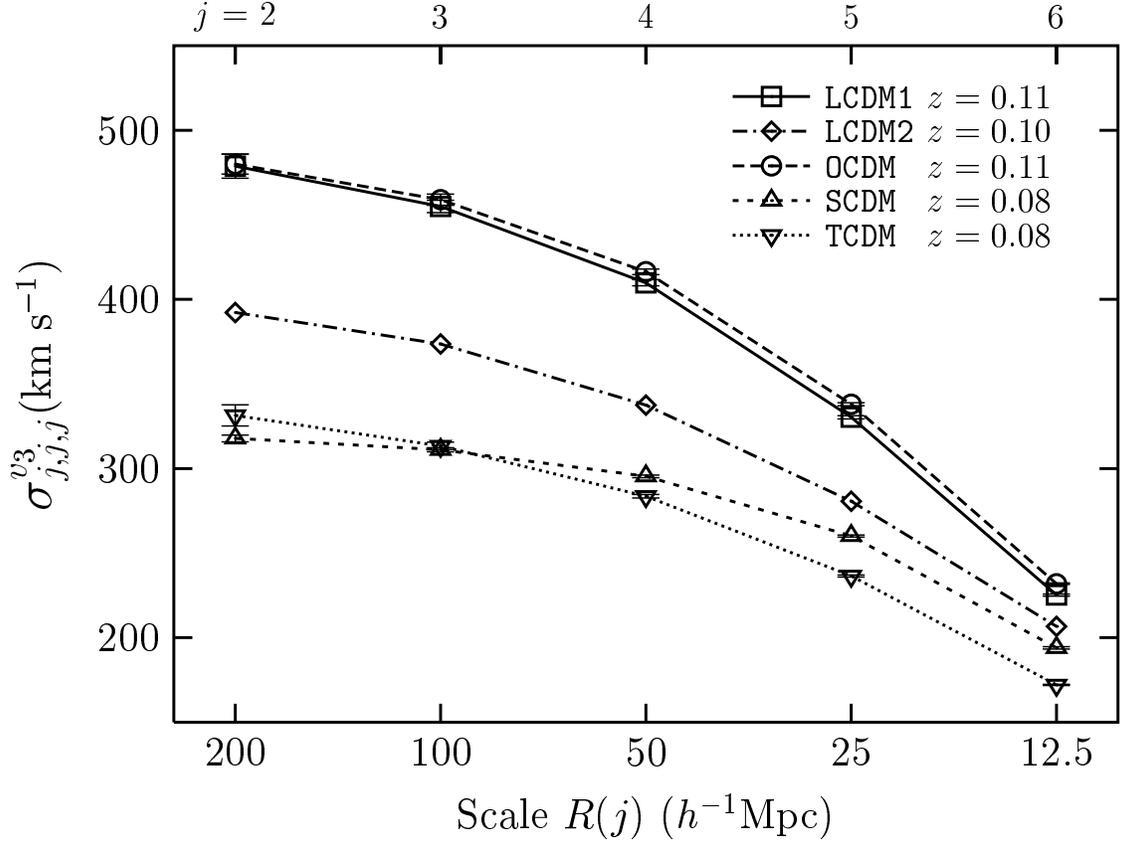}
\caption[f3.eps] {The real-space velocity dispersion as
defined in equation (42). Only the ${\bf j}$-diagonal
modes, i.e. the modes in cubic cells, are plotted. The units of
velocity is physical. The scale in 1-D is
$R(j) = 2^{-j}L$, where $L=800$h$^{-1}$Mpc for all models. This
definition is assumed in all the figures.
The result of LCDM2 model is always plotted without error bars here and
below,
since there is only one realization for this model. }
\label{Fig3}
\end{figure}

\begin{figure}
\epsscale{0.9}
\plotone{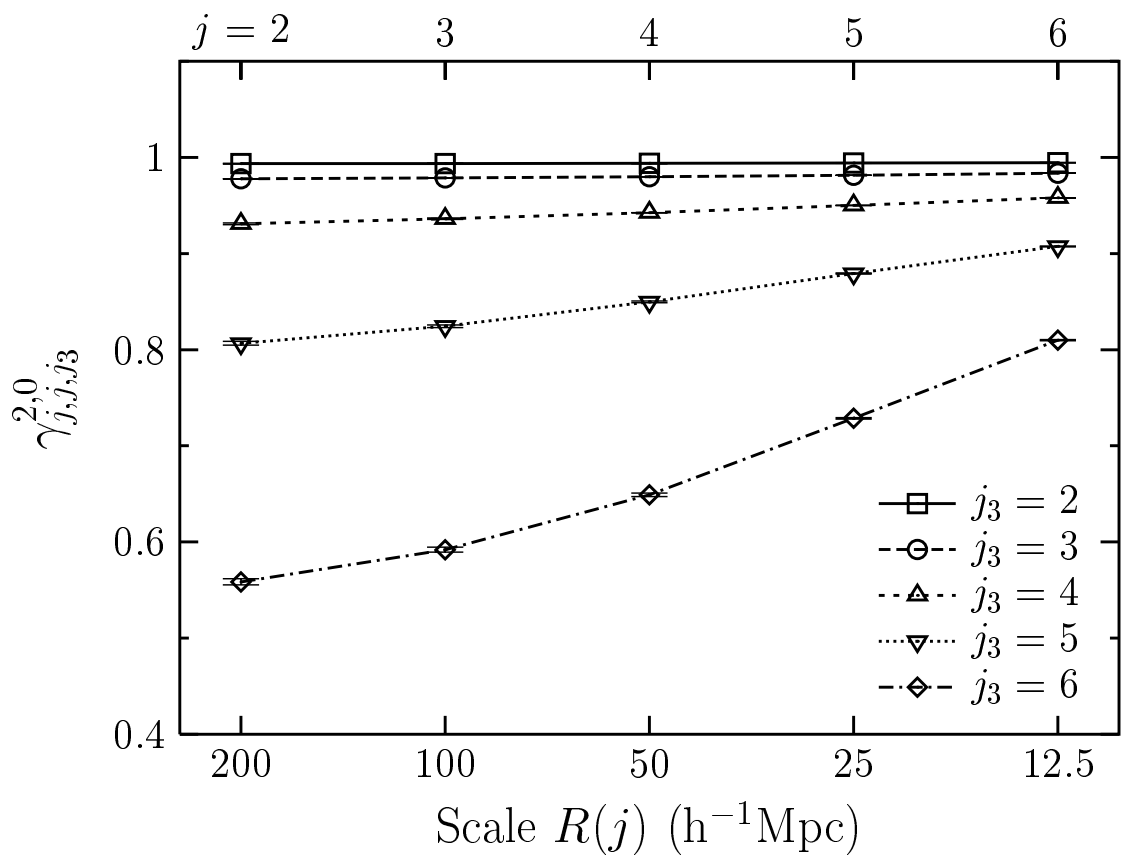}
\caption[f4.eps] {The factor of the non-linear redshift distortion
$\gamma^{2,0}_{\bf j}$ in equation (74). It is almost
independent
of $j_\perp$ when the LOS scale is above 50 h$^{-1}$Mpc.
} \label{Fig4}
\end{figure}

\begin{figure}
\epsscale{0.9}
\plotone{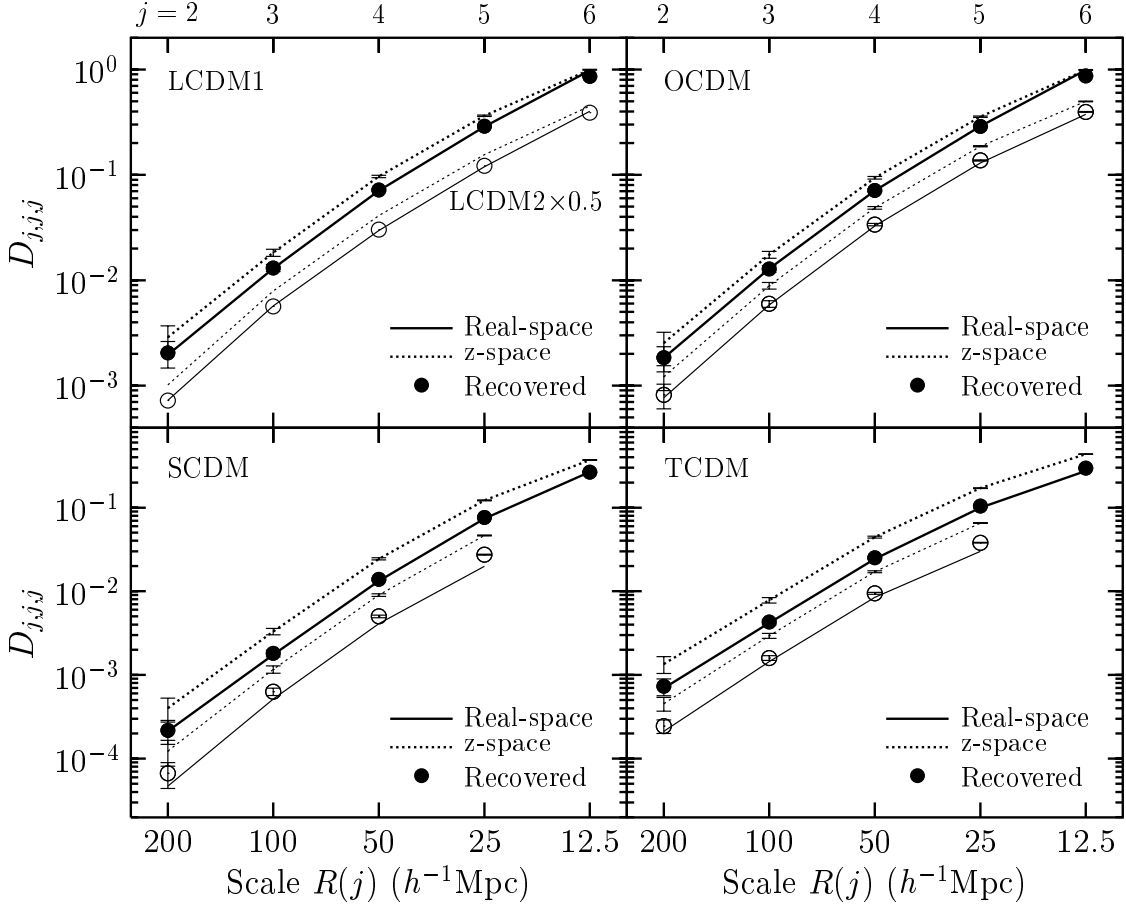}
\caption[f5.eps] {The ${\bf j}$-diagonal DWT 2nd moment $D_{\bf j}$
recovered using equation (75). The
thick lines are at $z=0.11$, for the LCDM1 and the OCDM models, and 0.08 for
the SCDM and the TCDM models. The thin (lower) line in the upper left panel
is the LCDM2 model at $z = 0.10$ and multiplied by 0.5. The other thin
lines are at $z=1.13$ for the OCDM model, and 0.71 for the SCDM and
the TCDM models. The redshift quoted here is the 
redshift at the center of each light-cone output. For clarity, the
real-space $D_{j,j,j}$ is plotted without symbols or error bars, and
the redshift distorted $D_{j,j,j}$ is plotted without symbols. These
treatments also apply to Figures 7 and 8. The SCDM and TCDM models have
significantly weaker initial power at small wavelengths, so the Poisson
noise
in the simulations is dominant on small scales at high redshifts. For this
reason, the 2nd moment $D_{6,6,6}$ is not shown
for the SCDM and the TCDM models at $z=0.71$.
} \label{Fig5}
\end{figure}

\begin{figure}
\epsscale{0.6}
\plotone{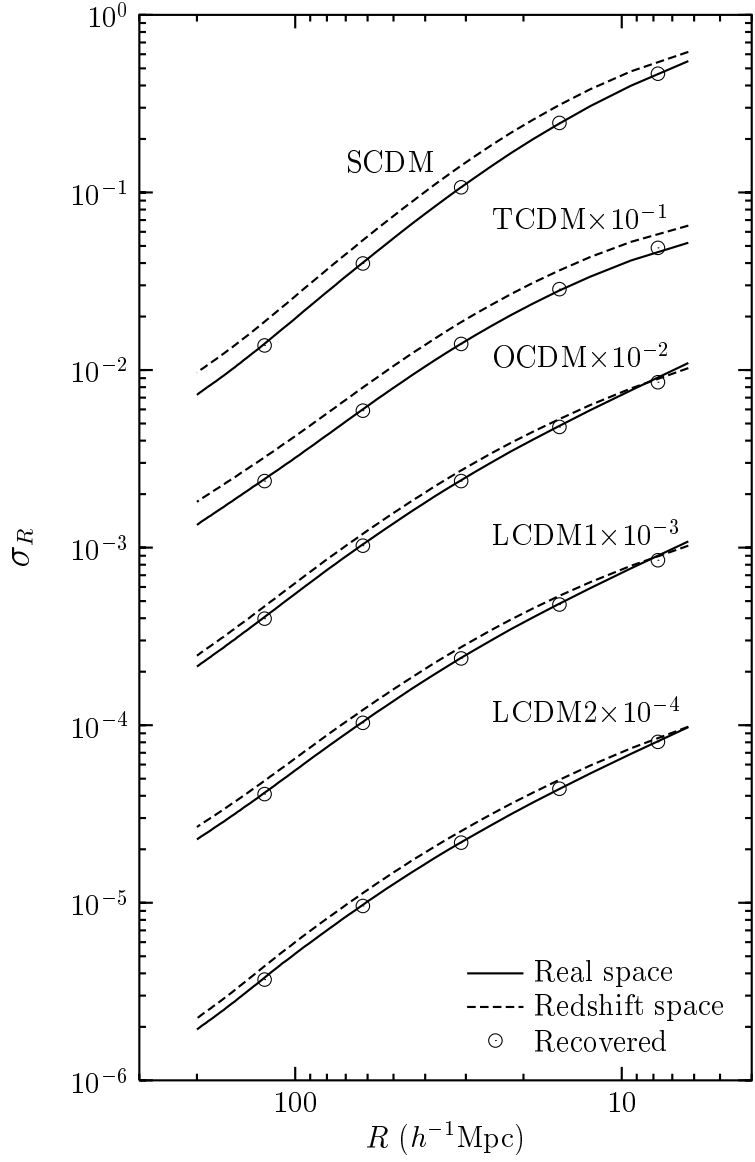}
\caption[f6.eps] {The recovery of $\sigma_R$, the rms density fluctuation
in a sphere of radius $R$, via $\sigma_R^S \simeq
[\gamma^{2,0}_{j,j,j}(1 + 2/3\beta + 1/5\beta^2)]^{1/2} \sigma_R$.
The models are shifted for easy reading.
The recovery is performed where the volume of the DWT
cell conveniently matches that of a sphere of radius $R$.
} \label{Fig6}
\end{figure}

\begin{figure}
\epsscale{0.9}
\plotone{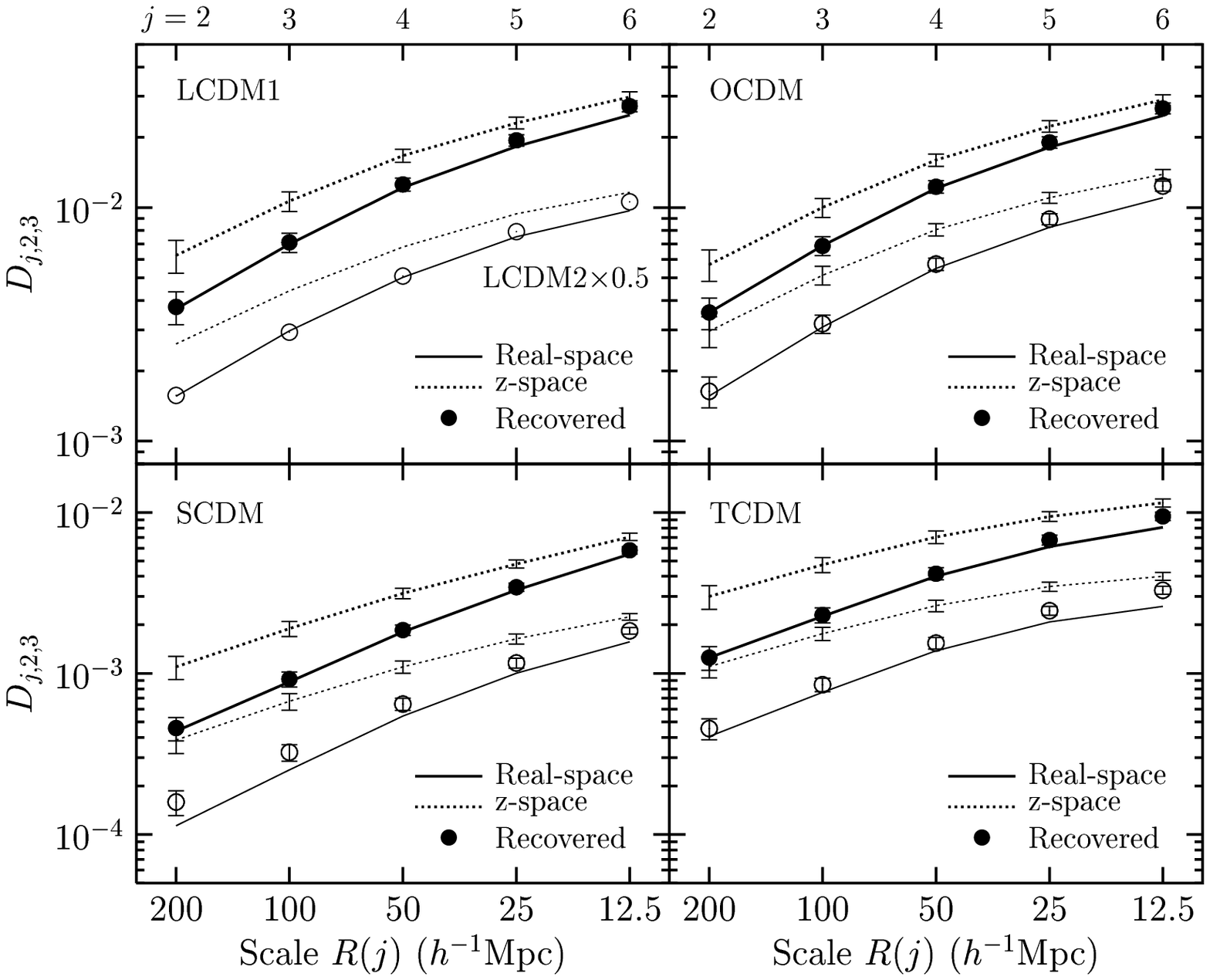}
\caption[f7.eps] {Same as Figure 5, except that this is
for off-diagonal modes $D_{j,2,3}$.
} \label{Fig7}
\end{figure}

\begin{figure}
\epsscale{0.9}
\plotone{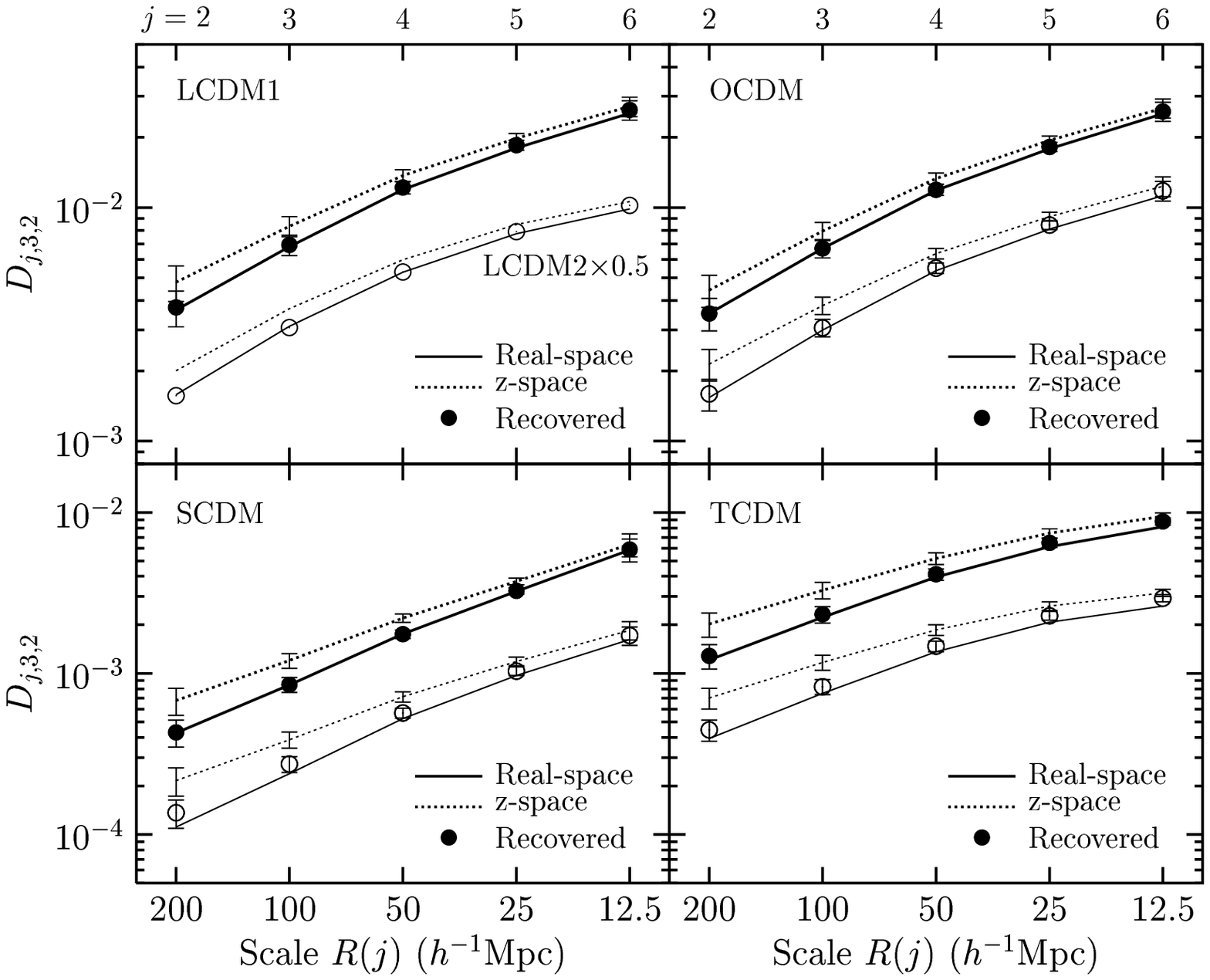}
\caption[f8.eps] {Same as Figure 5. except that this is
for off-diagonal modes $D_{j,3,2}$.
} \label{Fig8}
\end{figure}

\begin{figure}
\epsscale{0.6}
\plotone{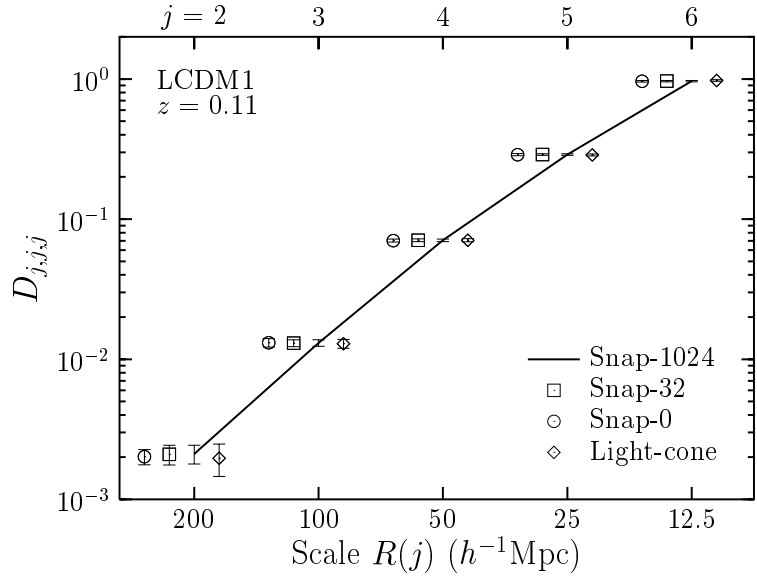}
\caption[f9.eps] {Snap-$n$, $n = 0, 32, 1024$, is calculated from 
the snapshot output with $n$ random shifts in grid.
Snap-0, Snap-32, and the light-cone data are displaced horizontally 
with respect to Snap-1024 data for easy identification.
} \label{Fig9}
\end{figure}

\begin{figure}
\epsscale{0.6}
\plotone{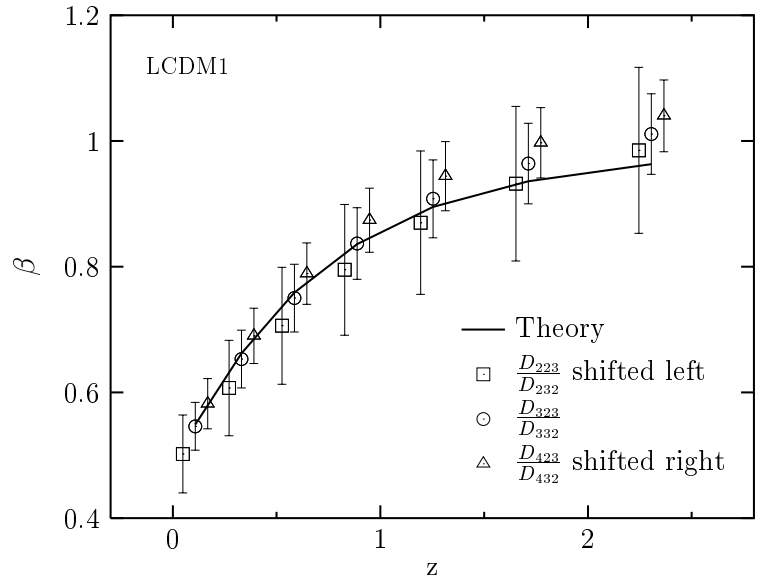}
\caption[f10.eps] {The $\beta$ parameter estimated by estimator eq.(81) for the
LCDM1 model.
} \label{Fig10}
\end{figure}

\begin{figure}
\epsscale{0.6}
\plotone{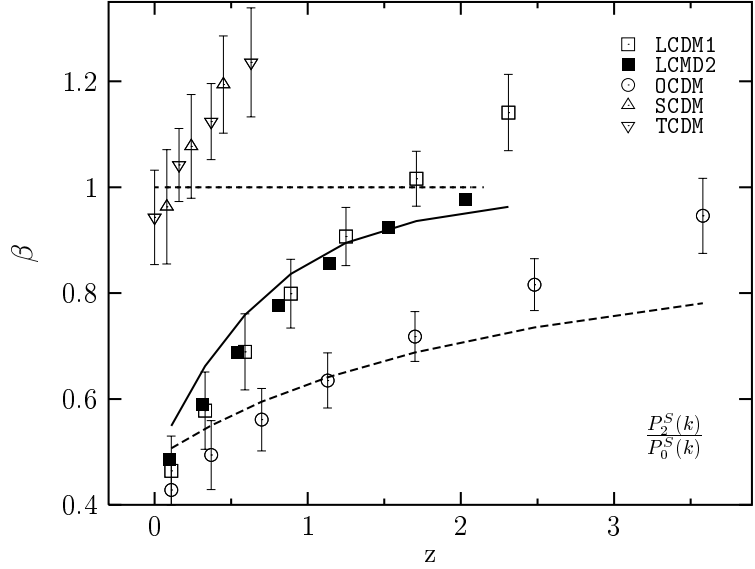}
\caption[f11.eps] {The $\beta$ parameter estimated by quadrupole-to-monopole
ratio. The TCDM model is shifted to the left for clear identification. The lines
are theoretical $\beta$ from equation (69) for each model.
} \label{Fig11}
\end{figure}

\begin{figure}
\epsscale{1}
\plotone{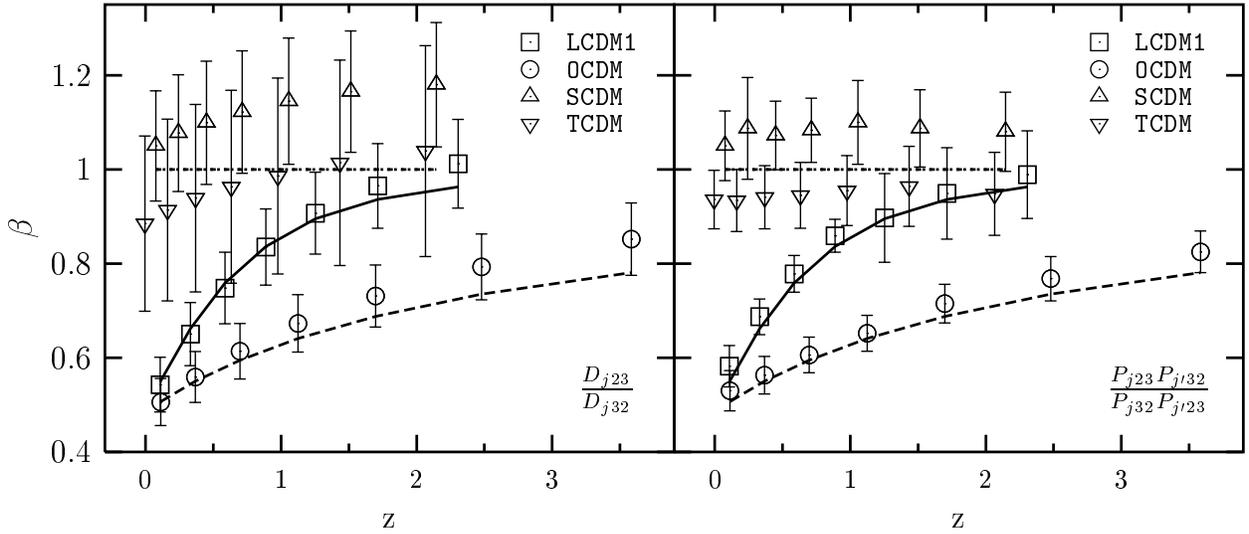}
\caption[f12.eps] {The left panel is the $\beta$ parameter given by an average
over that estimated
by eq.(81) with modes $j=2,3,4$ for the four models, and the right panel is
the $\beta$ parameter estimated by eq.(84).
The TCDM model is shifted to the left for clear identification. The lines
are theoretical $\beta$ from equation (69) for each model.
} \label{Fig12}
\end{figure}

\end{document}